%% file: SIDM.tex
\documentclass[useAMS,usenatbib,twocolumn]{mn2e}
\usepackage{natbib,graphics,epsfig}
\usepackage{times}
\usepackage{amsmath}
\usepackage{amssymb}
\usepackage{textcomp}

\usepackage{graphicx}
\usepackage{color}
\input{rgb}

\newcommand{\kms}{km~s$^{-1}$}
\newcommand{\Msun}{M$_{\odot}$}

\title[Dwarf Galaxies in SIDM vs CDM]{All about baryons: revisiting SIDM predictions at small halo masses}
\author[Bastidas-Fry et al.]{A.Bastidas Fry$^{1}$,
         F.Governato$^{1}$\thanks{E-mail:(FG); fabiog@astro.washington.edu},
         A.Pontzen$^{2}$,
         T.Quinn$^{1}$,
         M.Tremmel$^1$,
         L.Anderson$^1$,
 \newauthor
         H.Menon$^3$,
        A.\,M.Brooks$^{4}$ and
        J.Wadsley$^{5}$
\\
$^{1}$Astronomy Department, University of Washington, Box 351580, Seattle, WA, 98195-1580 \\
$^{2}$ UCL, Department of Physics \& Astronomy, Gower Place, London WC1E 6BT, UK\\
$^{3}$ Department of Computer Science, University of Illinois at Urbana-Champaign, USA\\
$^{4}$Dept. of Physics \& Astronomy Rutgers Univ. 136 Frelinghuysen Rd, Piscataway, NJ 08854  \\
$^{5}$Dept. of Physics and Astronomy, McMaster Univ., Hamilton, Ontario, L88 4M1, Canada
}

\begin{document}
\date{Revised version April 2015; In original form Dec 2014}

\pagerange{\pageref{firstpage}--\pageref{lastpage}} \pubyear{2012}

\maketitle

\label{firstpage}

\begin{abstract}
  We use cosmological hydrodynamic simulations to consistently compare
  the assembly of dwarf galaxies in both $\Lambda$ dominated, Cold
  (CDM) and Self--Interacting (SIDM) dark matter models. The SIDM
  model adopts a constant cross section of 2 $cm^{2}/g$, a relatively
  large value to maximize its effects.  These are the first SIDM
  simulations that are combined with a description of stellar feedback
  that naturally drives potential fluctuations able to create dark
  matter cores.  Remarkably, SIDM fails to significantly lower the
  central dark matter density within the central 500pc at halo
  peak velocities V$_{max}$ $<$ 30 \kms. This is due to the fact that
  the central regions of very low--mass field halos have relatively
  low central velocity dispersion and densities, leading to time
  scales for SIDM collisions greater than a Hubble time. CDM halos
  with V$_{max}$ $<$ 30 \kms \,have inefficient star formation, and
  hence weak supernova feedback.   At a fixed 2 $cm^{2}/g$ SIDM
    cross section, the DM content of very low mass CDM and SIDM halos
    differs by no more than a factor of two within 100-200pc.  At
  larger halo masses ($\sim$ 10$^{10}$ \Msun), the introduction of
  baryonic processes creates field dwarf galaxies with dark matter
  cores and central DM$+$baryon distributions that are effectively
  indistinguishable between CDM and SIDM.  Both models are in broad
  agreement with observed Local Group field galaxies across the range
  of masses explored.  To significantly differentiate SIDM from CDM at
  the scale of  faint dwarf galaxies, a velocity dependent cross
  section that rapidly increases to values larger than 2 $cm^{2}/g$
  for halos with V$_{max}$ $<$ 25-30 \kms needs to be introduced.
 \end{abstract}


\section{Introduction}

In this paper we use high resolution SPH + N-Body cosmological
simulations to focus on two main questions: Can we identify a clear
difference between Cold Dark Matter and Self--Interacting Dark Matter
(CDM and SIDM respectively) predictions for the structural and
observable properties of dwarf galaxies? How do the properties of CDM
and SIDM halos differ at halo masses below 10$^{10}$ \Msun, where star
formation (SF) becomes very inefficient and the effect of the
underlying DM component should dominate?

SIDM was originally introduced as a solution for the so called
``core-cusp'' problem, the excess of dark matter predicted by the CDM
model at the center of field and satellite dwarf galaxies compared to
observations \citep{moore98b,oh08,walker11,adams14}.  In the SIDM
model dark matter particle collisions isotropize the cores of
galaxies and transfer mass outward from  the dense central
regions of DM halos over cosmic time scales.  This process creates
large cores, more spherical halos, and a signature flat radial profile
of the DM velocity dispersion \citep{spergel00,burkert00}.  A SIDM
model with fixed cross section and elastic collisions represents the
simplest model in a large class of plausible ``dark sector'' DM
models.  However, significantly more complex interactions are
possible, including Yukawa potentials \citep{feng09,feng10,loeb11}, and
cooling or atomic dark matter \citep{cyr-racine13,schutz15,buckley14}.
The dynamics of SIDM was first implemented in cosmological simulations
as a fluid \citep{moore00} and then as elastic collisions between
particles \citep{spergel00,burkert00,
  dave01,colin02,strigari07,martinez09, koda11,vogelsberger12}.

Numerical studies combined with observations (the ellipticity of
galaxy clusters, the abundance of and the core size of field dwarf
galaxies) have constrained the cross section of SIDM to be of the
order of $\sim$ 0.1 --1 cm$^2/g$ \citep{loeb11,rocha13,peter2013,
  zavala13}.  However, a variable SIDM cross section (a decreasing
function of the DM particles' relative velocity) can preserve the
ellipticity and density of galaxy clusters while producing cores in
dwarf galaxies \citep[][but see \citet{newman13b} for evidence for DM
cores in clusters]{peter2013}.  While weakening existing constraints,
a variable SIDM cross section  is well motivated by ``hidden
  sector'' particle physics models \citep{zurek13}.  Further crucial
constraints on large SIDM cross sections \citep{elbert14} come from
the necessity of forming cores in very faint galaxies without
evaporating the satellites of MW-like halos and galaxy sized halos in
clusters as they interact with a dense DM environment
\citep{gnedinSIDM,vogelsberger12}.

Unfortunately, a large fraction of the astrophysically driven support
for non-standard DM models has come so far from simplified simulations
that lack the complexities of ``baryon physics'' and follow only the
the assembly of the DM component \citep[see the review by ][]{brooks14r}.
The necessity to couple a DM model with baryon physics comes from the
existence of ``bulgeless galaxies'', a problem that requires the
removal of low angular momentum gas from galaxies
\citep{binney01,vdbosch01a,G10} through feedback processes \citep{brook11}.
Further motivation for including baryon physics includes the necessity
to quench SF in galaxy satellites \citep{Klypin99, Moore99,donghia03},
especially if the SIDM power spectrum is similar to CDM at small
scales and subhalos survive evaporation \citep[as found
by][]{zavala13}.

Crucially, analytical and numerical work have shown that feedback
lowers the central DM density in galaxies, creating gas outflows and
repeated fluctuations in the gravitational potential
\citep{mashchenko08,
  delpopolo09,G10,PG12,dicintio14,teyssier13,velliscig14}.  This
results in irreversible energy transfer to the DM \citep[see also
review by][]{review14}. These outflows, generated by a bursty SF,
  have strong observational support
  \citep{vanderwel11,martin12,lundgren12,dominguez2014,geach15}. Bursty
  SFH have also been robustly associated with the build up of the
  stellar content of galaxies in the 10$^8$--10$^{10}$ \Msun range
  \citep{tolstoy09,kauffmann14}. The ability of feedback to
dynamically heat the DM can potentially remove the need for SIDM at
galactic scales. As a result it is fundamental to identify the unique
differences between CDM and SIDM when both models include an explicit
treatment of the physics of galaxy formation.

This study is able to follow the evolution of field SIDM halos at
masses below $\sim$10$^{10}$ \Msun\, (V$_{max}$\footnote{where
  V$_{max}$ is defined as the peak of the rotation curve, with
  V$_{rot} = (GM/r)^{1/2}$.} $<$ 40 \kms ).  This is an important
regime as (1) observational data are becoming robust
\citep[e.g.,][]{papastergis15}, and (2) feedback processes become less
efficient with declining V$_{max}$
\citep{G12,penarrubia12,dicintio14}. The most recent simulations
  \citep{madau14a,GWDM,onorbe15} show core formation when only
  10$^{5.7-6}$ \Msun of stars have formed.  Also, it is often naively
assumed that SIDM will form DM cores at all dwarf masses. However, at
these small halo masses the trend of later assembly epoch of the first
progenitor \citep{li08}, and consequentially lower central densities
\citep{avila05}, will affect the timescale and extent of SIDM core
formation in field galaxies beyond simple scaling calculations.
Furthermore, the relative velocity of a halo and the surrounding DM
background is crucial in determining the collision rate, boosting it
in small halos (that have intrinsic low $\sigma$).  Our subset of
DM-only simulations is also one of the first to compare very small
field and satellite galaxies at similar mass and spatial
resolution. The central density of field and satellite dwarfs (that
move fast through a dense DM environment) may then be significantly
different, but it has not been properly compared yet.

In this work we adopt a constant velocity cross section of 2 cm$^2$/g
in the SIDM runs (a relatively large value to maximize its
  effects) and a common description of SF and feedback in both CDM
and SIDM models. The SF and feedback prescriptions have been shown to
form CDM galaxies with SF efficiency, photometric, and kinematic
properties close to those of real ones
\citep{oh11,munshi13,christensen14,shen14, brooks14}.  These
simulations are the first high resolution simulations to compare the
assembly of dwarf galaxies in CDM vs SIDM cosmologies including a
description of SF and supernova feedback that creates realistic
galaxies while creating DM cores through ``DM dynamical heating.''
Other recent numerical work \citep{vogelsberger14} focused on small
galaxies in SIDM and explored the role played by a variable cross
section, however, the feedback recipe implemented in their work does
not generate DM cores.

 In \S 2 we describe the simulation and the
code used. In \S 3 we present the results, which are then discussed in
\S 4.

\section{Methodology}\label{methods}

\subsection{ChaNGA, star formation  and bursty feedback  with outflows}

The simulations were run in a full cosmological context and to a
redshift of zero using the N-body Treecode $+$ Smoothed Particle
Hydrodynamics (SPH) code {\sc ChaNGa}
\citep{changa08,changa13,menon14}\footnote{ {\sc ChaNGa} is part of
  the {\sc{AGORA}} group, a research collaboration with the goal of
  compare the implementation of hydrodynamics in cosmological codes
  \citep{AGORA}.  {\sc ChaNGa} is available here:
  http://www-hpcc.astro.washington.edu/tools/changa.html}. {\sc
  ChaNGa} includes standard physics modules previously used in {\sc
  Gasoline} \citep{wadsley04,wadsley08,stinson12} including a
treatment of metal line cooling, self shielding, cosmic UV background,
star formation, ``blastwave'' SN feedback and thermal feedback from
young stars \citep{stinson06,stinson12}.  The SPH implementation
includes thermal diffusion \citep{shen10} and eliminates artificial
gas surface tension through the use of a geometric mean density in the
SPH force expression \citep{ritchie01,GWDM,menon14}.  This update
better simulates shearing flows with Kelvin-Helmholtz
instabilities. For consistency with our previous work comparing CDM
and WDM scenarios, we adopted the same feedback and SF parameters as
in \citet{GWDM}.  A Kroupa IMF is assumed \citep{kroupa2001} and the
density threshold for SF is set at 100 amu/cm$^3$. Limiting star
formation to dense gas regions is a realistic approach and
concentrates feedback energy \citep{brook12,christensen14b,agertz14}.
100\% of SN energy is coupled to the surrounding gas.

\begin{table*}
\begin{tabular}{ccccc}
\hline
Name               & Physics         &  DM/gas particle mass \Msun   &  Force Softening (pc) & halo mass range (\Msun)  \\
\hline
\hline
h516CDM        & SF and DM-only         & 1.6 10$^4$/3.3 10$^3$   &   86      & 10$^9$ -  5 $\times$ 10$^{10}$             \\  \hline
h516SIDM        & SF and DM-only        & 1.6 10$^4$/3.3 10$^3$   &  86       & 10$^9$ -  5 $\times$ 10$^{10}$            \\  \hline
h2003CDM        & SF and DM-only        & 0.67 10$^4$/1.4 10$^3$  & 64        & 4 $\times$ 10$^8$ - 1  $\times$ 10$^{10}$             \\  \hline
h2003SIDM        & SF and DM-only       & 0.67 10$^4$/1.4 10$^3$  & 64        & 4 $\times$ 10$^8$- 1  $\times$ 10$^{10}$            \\ \hline
40Thieves-CDM   &   DM-only         &   0.81  10$^4$              & 64       &  4 $\times$ 10$^8$- 2.7 $\times$ 10$^{10}$            \\ \hline
40Thieves-SIDM   &   DM-only        &   0.81  10$^4$              & 64       &  4 $\times$ 10$^8$- 2.7 $\times$ 10$^{10}$              \\ \hline
40Thieves-SIDM.hr  &   DM-only      &   0.24  10$^4$              & 64       & 1.25  $\times$ 10$^8$  - 2.7 $\times$ 10$^{10}$             \\ \hline
h148CDM             &   DM-only     &   1.93  10$^4$                   & 86       &  10$^9$- 2 $\times$ 10$^{12}$ \\ \hline
h148SIDM            &   DM-only     &    1.93  10$^4$                  & 86       & 10$^9$- 2 $\times$ 10$^{12}$ \\ \hline
\end{tabular}
\caption{{\sc Simulations and   physics parameters explored in our simulations}. All SIDM runs adopt $\sigma_{SIDM} = 2 [{\rm cm}^2\,{\rm g}^{-1}].$ The mass range shows
  halos with at least 50,000 DM particles within their virial radius. The main halos in each volume are studied with several million particles each. The DM-only subsample is one of the first to simulate a sample of  field and satellite halos at similar mass and spatial resolution. This approach allowed us to study the effect of the  environment on the DM distribution of the halos in the two populations. }
\label{table:ref_points}
\end{table*}

\subsection{SIDM Implementation and Analytical Expectations}

The SIDM implementation closely follows the standard Monte Carlo
approach described in previous works \citep{dave01,vogelsberger12},
and here is only briefly summarized, while tests are presented in
Appendix A. SIDM interactions are modeled under the assumption that
each simulated DM particle represents a patch of DM phase-space
density and that the probability of collisions is derived from the
collision term in the Boltzmann equation. Collisions are then elastic
and explicitly conserve energy and momentum. For a detailed
discussion see \citep{rocha13} and also
\citep{yoshida00,donghia03,kaplinghatSIDM}.  The number of
interactions N that occur in a region with local DM density $\rho$ in a
time $\Delta t$ is

\begin{equation}
N= \rho  v\sigma_{dm} \Delta t
\end{equation}

In practice to simulate interactions for discrete N-body DM particles
we use equation one as the implied probability for a particle to
scatter. This assumption is valid as $\Delta t$ approaches zero or
numerically where $\Delta t$ is chosen to be much smaller than needed
to avoid multiple collisions. At each time step our numerical code
calculates the relative velocity and density $\rho$ using the same
functional SPH kernel as in hydro calculations of each DM particle in
relation to its 32 nearest neighbors. These values are used to calculate
the probability that the DM particle interacts with one of its
neighbors. Similar to Vogelsberger et al. (2012), a scattering event
may occur at each time step between particles i and j if a uniform
random variable in the interval (0,1) is less than P$_{ij}/2$ (the
probability is divided by two in order to account for the fact that
each set of particles are compared to each other twice).  While
producing similar results, this approach differs slightly from the one
adopted in \cite{rocha13}, where collisions are defined between DM
particle pairs and the probability of interaction P(i$|$j) is
explicitly = P(j$|$i).  When a particle collision is detected we
isotropically and elastically scatter the particles to random angles.

The interaction cross section $\sigma$ for all SIDM runs in this work was set to
the relatively large value of 2 g/cm$^2$, close to the upper limit
for constant $\sigma$, in order to maximize its effects compared to
that of baryon physics.  The way in which SIDM reshapes
different halos can be readily understood
setting $N=1$ in eq.1 and
$v=v_{\mathrm{max}}$ (where $v_{\mathrm{max}}=\max \sqrt{GM(<r)/r}$) to obtain a characteristic
maximum timescale on which a given density is stable,
\begin{equation}
\tau_{\mathrm{SI}} = \frac{1}{\rho v_{\mathrm{max}} \sigma_{dm}}\textrm{.}
\end{equation}
After a collision, the change in momentum is usually enough to fully
eject the particle \citep{kahlhoefer14}, so dark matter densities must
drop over a few $\tau_{\mathrm{SI}}$.

One important conclusion to draw from eq.3 is that not every SIDM
model will necessarily form significant cores in the smallest
halos. {\it Low halo peak velocities and low central densities
  (at a fixed radius) may result in $\tau_{\mathrm{SI}}$ comparable to
  the lifetime of a halo and therefore in the preservation of cusps.}  Another
important consequence is that in SIDM the central densities of very
small satellite halos could be differentiated from those of their field
counterparts by the interaction with the DM halo of a more massive
host as the orbital velocity is much higher than the internal
velocities of the satellite.  We test these analytical predictions in
the next section, where we present results of simulations that resolve
the internal structure of DM halos smaller than most previous SIDM
studies.  Observational evidence of DM cores in small field halos
where baryonic processes may be inefficient (V$_{max}$ $<$ 20-30 \kms
and in satellite galaxies (which move at 100-200 \kms in a $\sim$
200-1000 $\times$ $\rho_{crit}$ DM field) can then provide useful
lower limits to the SIDM cross section.

\section{CDM and SIDM Simulations}
\label{sims}

In all simulations we assumed a $\Lambda$ dominated cosmology ($\Omega_0$
$=$ 0.26 $\Lambda$ $=$ 0.76, $\sigma_8$ $=$ 0.77, n$=$0.96) and used
the ``zoom-in'' technique to achieve high resolution in the regions of
interest \citep{katz93}. The gravitational force spline softening
length is in the range 64-86~pc \citep{power03} and the smoothing
length for the gas component is allowed to shrink to 0.1 the force
softening.  Simulations start at $z = 120-100$.  The combination of
simulations (see Table 1)  allow us to compare the effects of SIDM
and baryon physics on a range covering almost four orders of magnitude in
halo masses and a variety of environments from the field to the dense
region within the virial radius of a large field disk galaxy.

Cosmological simulations of well resolved halos with mass $<$
10$^{10}$ \Msun ~(corresponding to a halo peak velocity V$_{max} <$ 30
\kms) are particularly relevant for SIDM, as baryonic effects at these
scales should be limited \citep{G12,penarrubia12}. Based on equation (2), the
ability of SIDM to lower the central density of halos depends on the
DM $\rho$ and the typical particle velocity, which is lower in less
massive halos.  Interestingly, previous CDM simulations have suggested
that the central density of small mass field halos {\it decreases}
with their halo mass \citep{avila05,li08}.
In a SIDM scenario, extending this trend to very small halos
could prevent the formation of low density central cores by increasing
the time scale of DM-DM collisions.

\begin{figure}
\hspace{-1cm}\includegraphics[width=0.55\textwidth]{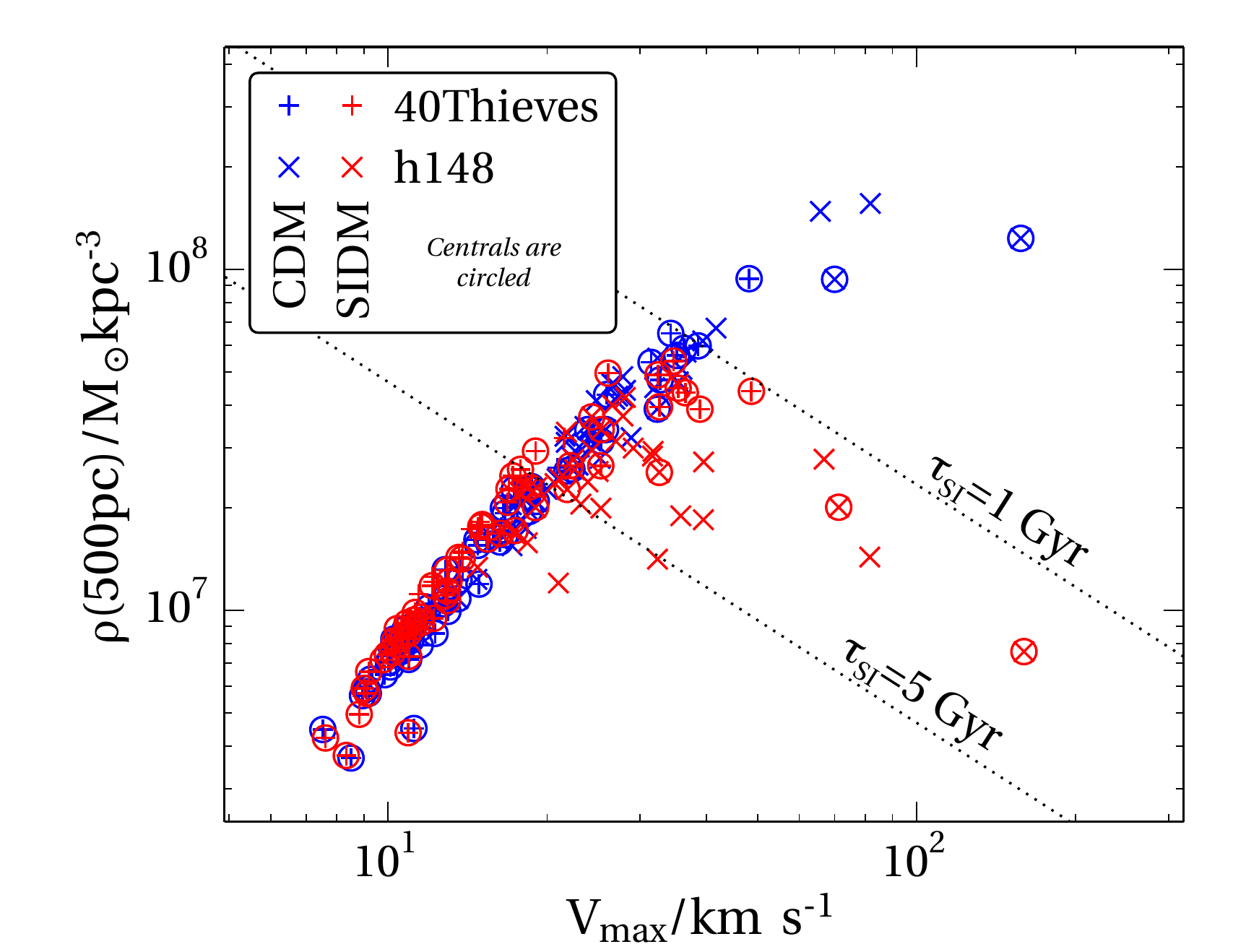}
\caption{{\sc DM-only Simulations:}~The DM density measured at 500 pc
  as a function of halo maximum velocity and dark matter model. Blue
  crosses:~CDM, red crosses:~SIDM. Circled symbols correspond to
  isolated halos, non-circled symbols show the satellites of the
  MW-sized, h148 halo.  At small halo masses the central densities of
  CDM and SIDM halos do not differ substantially, while at larger
  velocities/halo masses SIDM-only halos have lower central densities
  than their CDM-only counterparts.  This difference can be understood
  in terms of the dotted lines which show the maximum timescale
  $\tau_{\mathrm{SI}}$ at which collisions are significant in the SIDM
  case (see text for details). This result shows that fixed cross sections
  commonly adopted at the scale of larger systems (0.1-2 cm$^2$/g)
  would not be sufficient to form kpc sized cores in the smallest
  observable field halos with V$_{max}$ $<$ 30 \kms.  In the V$_{max}$
  30-60 \kms range SIDM satellites have central densities lower by a
  factor two compared to their field counterparts, due the added boost
  to SIDM collisions coming from satellites orbiting at high velocity
  in the dense DM halo of the host.  }
\label{fig1}
\end{figure}

\subsection{ DM-only Simulations: Environmental effects and cuspy halos at
  V$_{max}$ $<$ 30 \kms}

To study the structure of a significant sample of DM halos in a range
of environments we simulated four different regions with the zoomed-in
approach \citep{katz93}.  The first three regions (Table 1) are
centered on filamentary structures with a density close to average
measured inside a sphere of 5 Mpc in radius. The largest one of them
is nicknamed `The Forty Thieves' and includes several tens of halos
with V$_{max}$ $<$ 30 \kms, equivalent to a mass range
10$^8$-10$^{10}$\Msun, where the minimum mass refers to halos with at
least 50,000 DM particles within the virial radius. To complement this
dataset, a high resolution simulation of a massive halos and its
system of satellites has been included (h148). The halos in this
simulation have mass and spatial resolution similar to their `40
Thieves' counterparts and mass and spatial resolution better by factor two compared to
recent work \cite{vogelsberger12,zavala13}. In Appendix 1 we verified that our
results and the Monte Carlo implementation of SIDM collisions are not
affected by resolution effects.

\begin{figure}
\vspace{-4em}
\hspace{-0.4cm}\includegraphics[width=0.5\textwidth]{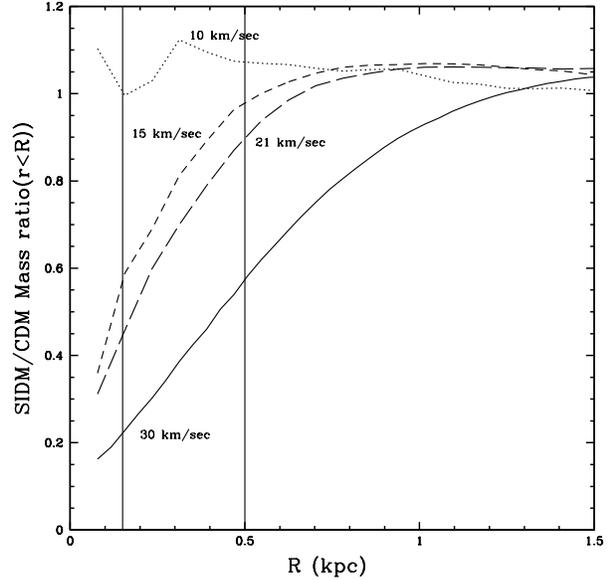}
\vspace{-10em}
\caption{{\sc The relative DM content as a function of radius in
    DM-only Simulations:}~The ratio of the cumulative DM content
  measured for a set of small halos over the 0-1.5kpc radial range
  (from the 40-Thieves run). Each CDM halo was matched with its 'twin'
  halo in the SIDM simulation. The spline kernel force resolution is
  64pc. Vertical lines mark 150 and 500 pc (2.3 and 5 softening
  lengths).  With our adopted cross section of 2 cm$^2$/g small CDM
  and SIDM become progressively difficult to differentiate, with DM
  content at a given radius being within a factor of two in the two
  cosmologies. Detecting such a difference in field dwarfs could
  require a next generation telescope (M.Walker, private
  communication). A SIDM with a variable cross section could show
a more significant difference compared to CDM. }
\label{figratio}
\end{figure}

\begin{figure*}
\centering{
\includegraphics[width=0.45\textwidth]{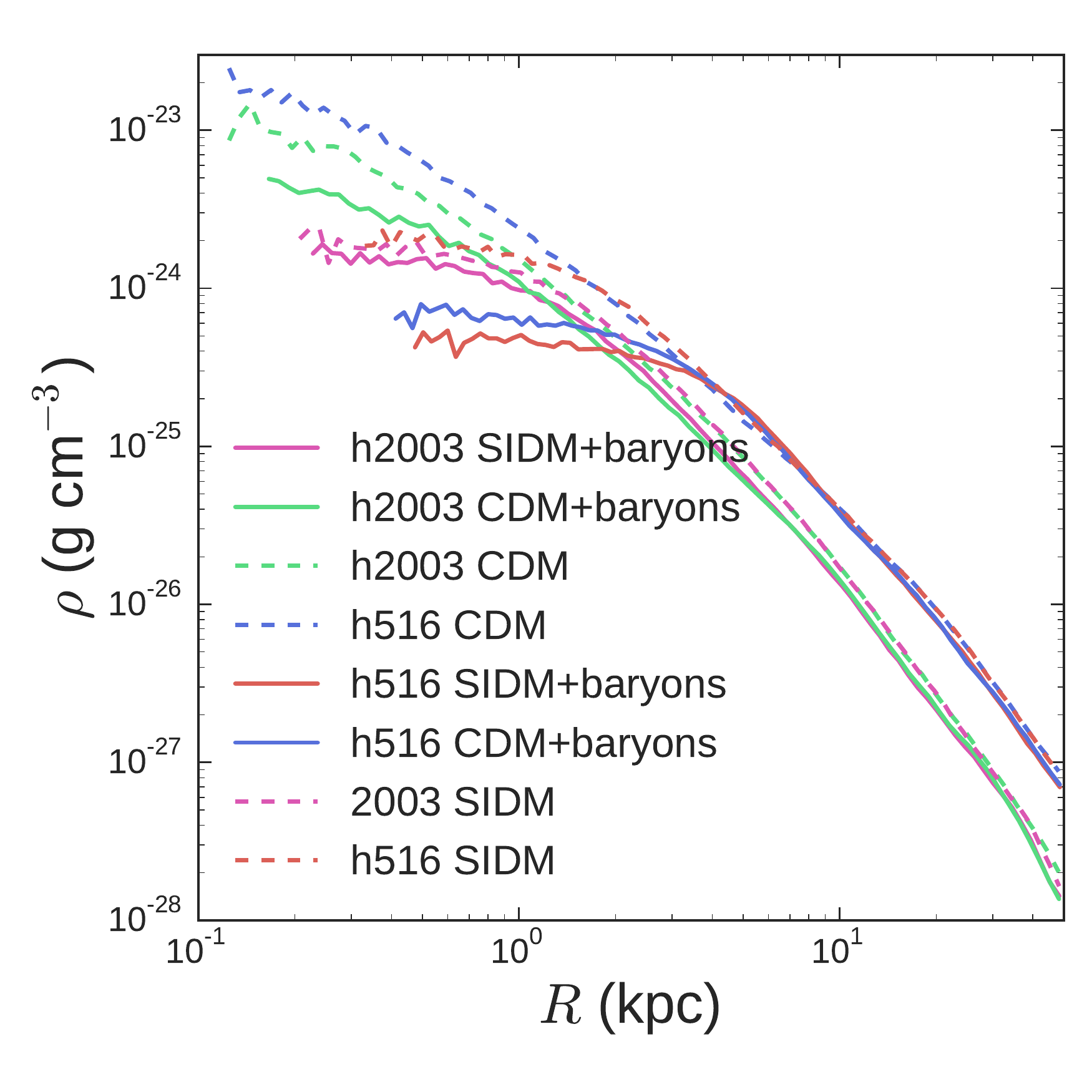}
\includegraphics[width=0.45\textwidth]{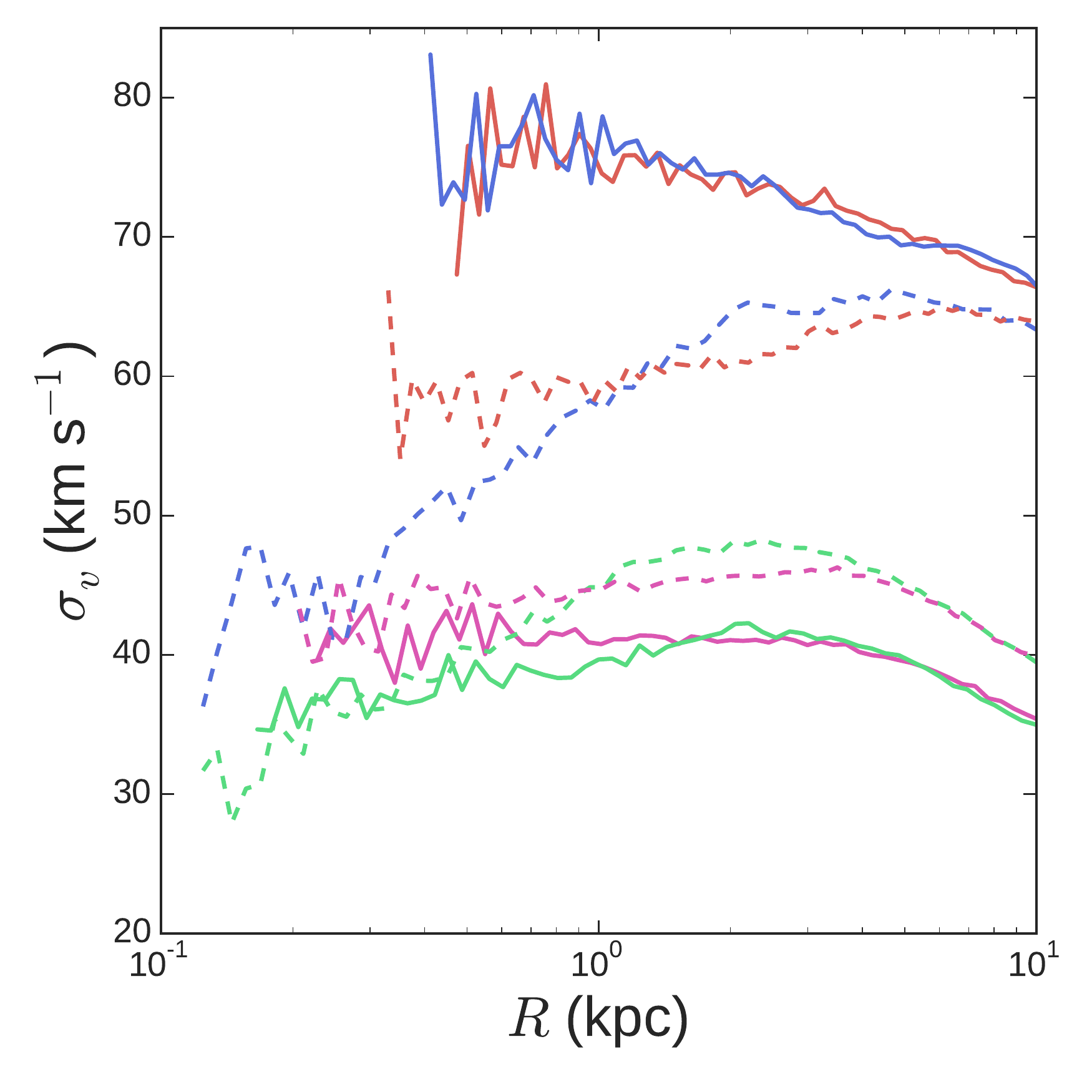}
}
\vspace{-0em}
\caption{{\sc Radial Profiles.}~Left:The radial density profile of the
  DM component for halos/galaxies h516 and h2003.  Right:The velocity
  dispersion of DM component in the same halos/galaxies. h516-CDM:
  blue. h516-SIDM: red. h2003-CDM: cyan. h2003-SIDM: magenta. Dashed:
  DM-only runs. Both density and dispersion profiles become similar in
  CDM vs SIDM once baryonic processes are introduced. }
\label{figprofile}
\end{figure*}

\begin{figure}
\vspace{-6em}
\hspace{-0.5cm}\includegraphics[width=0.5\textwidth]{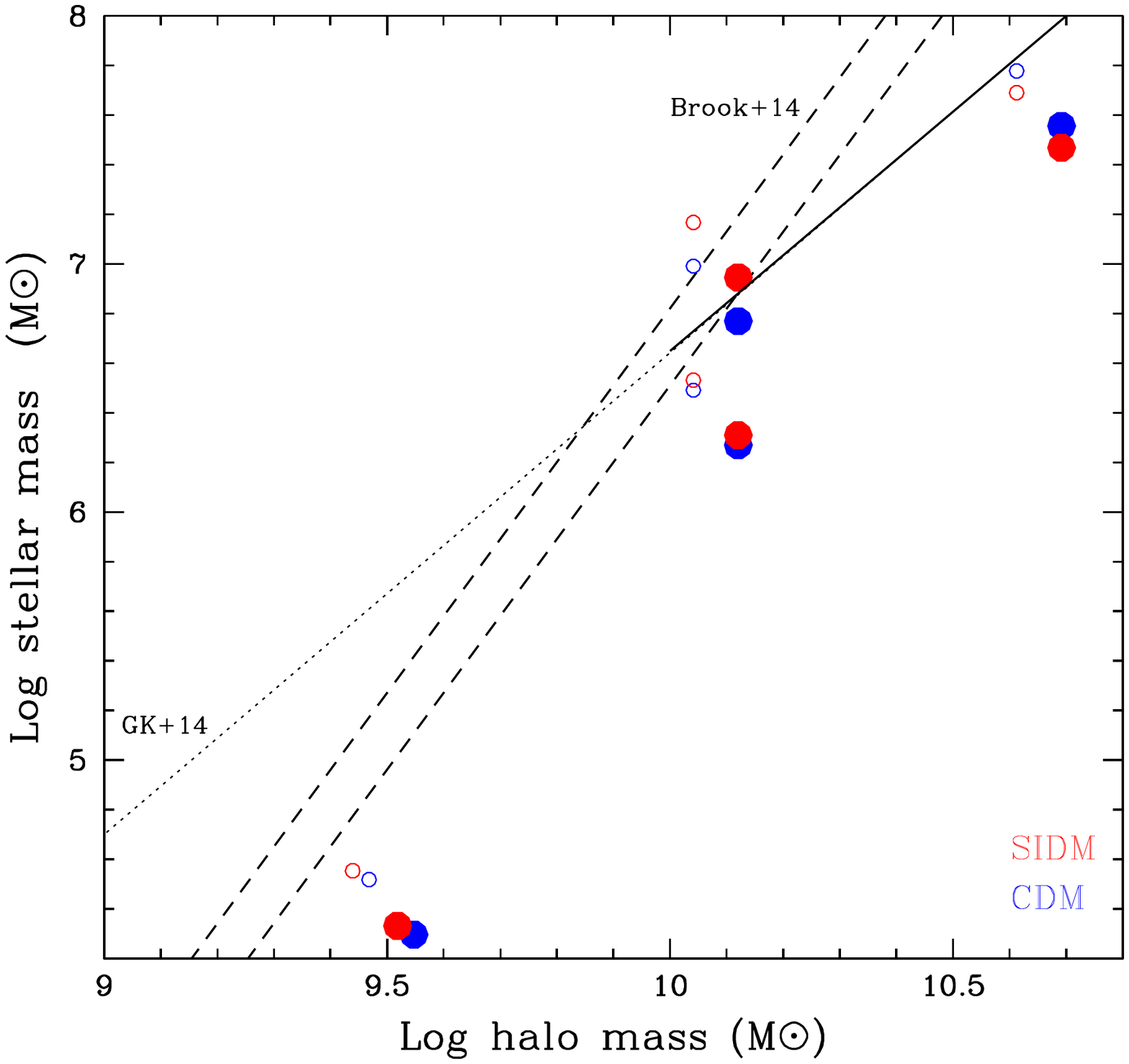}
\vspace{-8em}
\caption{{\sc Stellar Mass/Halo Mass relation:}~The Stellar Mass - Halo Mass relation
for the simulated galaxies. Dashed lines and solid lines  show the relation obtained from Local Group data \citep{brook14,gk14}
The relations are extrapolated below $\sim$  10$^{6.5}$ \Msun due to sample incompleteness (dotted lines). Circles show the raw data, solid dots show the simulation data correcting for
observational and simulation biases \citep{munshi13} in measuring stellar and halo masses. Overall the simulations produce the right amount of stars. The most massive halo converts about 1\% of gas into stars. The rapid drop in SF efficiency at halo masses below 10$^{10}$ \Msun is due  to the introduction of 'early feedback' (see text).
}
\label{figmass}
\end{figure}

In Figure \ref{fig1} we show the density measured at 500pc \citep[a
radius at which the rotation curves of many dwarf galaxies have been
resolved, e.g.,][]{oh11} for all resolved halos in our h148 ($\times$)
and ``the 40 Thieves'' ($+$) DM-only simulations as a function of
$V_{max}$.  We then over plot lines of fixed $\tau_{\mathrm{SI}}$ for
1 Gyr and 5 Gyr.  SIDM and CDM show identical central DM densities if
the typical scale for DM interactions is longer than 5 Gyr. At larger
peak velocities (and halo masses) with shorter timescales for
interaction, the SIDM densities decrease compared to their CDM
counterparts and fall in a valley between the 1 and 5 Gyr lines,
matching results from previous works
\citep{rocha13,vogelsberger12}. In other words, SIDM interactions
force the DM density to drop until, going to smaller and smaller halo
masses, the interaction timescale rises to a significant fraction of
the Hubble time and SIDM (with a $\sigma$ = 2 cm$^2/g$) is not
effective anymore. Figure \ref{figratio} shows the SIDM vs CDM
  ratio of the enclosed mass as a function of radius for paired halos
  from the `40 Thieves' run. At radii currently tested by observations
  (100-500pc), but entirely neglecting baryonic physics, the
  difference between the two models decreases rapidly. A factor of two
  in the DM content within 100-500pc of faint field dwarfs, could be
  revealed by large spectroscopic samples of the associated stellar
  populations (M.Walker, private communication).  We also verified
  that at larger V$_{max}$ the density profiles of the DM-only halos
  match those recently published in \citep{elbert14}, which at
  V$_{max}$ $\sim$ 35 \kms form significant cores (see also next
  section where the effect of baryons in these larger systems is
  included).  This result confirms our simple analytical expectations
and shows that even with a significant constant SIDM cross
section, DM cores rapidly become smaller than our resolved scale
($\sim$ 100-200~pc or twice the spline kernel softening) in field
halos with virial mass $<$ 10$^{10}$ \Msun and V$_{max}$ $<$ 30 \kms.
Confirming the results from \cite{li08} and extending them to much
smaller halo masses and higher resolution we verified that the
increased $\tau_{\mathrm{SI}}$ at small halo masses comes not only
from a lower $\sigma_{DM}$ but also from lower cusp densities,
possibly due the later epoch of collapse of the central regions of a
halo. We verified the DM density within the central 250~pc decreases
by a factor of eight in CDM halos with mass from 10$^{8}$ \Msun
~compared to halos of 10$^{10}$ \Msun.  If we look at the average
densities as a function of V$_{max}$, another difference emerges
between the SIDM field (circled red crosses) and satellites (red
crosses), with satellite halos showing central densities lower by
about factor of two compared to field halos of similar V$_{max}$.
This important environmental difference could be due to the satellites
forming their central regions earlier, or to the significant boost to
$\rho$ and v$_{max}$ in eq.3 due to their orbiting 1) inside the dense
halo of a much more massive host and 2) at a much higher speed than
their internal velocity dispersion. Our simulations were able to
highlight this difference as our simulations resolve field {\it and}
satellites halos with significantly lower peak velocities (10-20 \kms)
than in previous works.

The main result from this section is that evidence of DM cores in real
galaxies with V$_{max} <$ 30 \kms ~would constrain the SIDM cross
section to values $\sigma$ $>>$ 2 cm$^2$/g when the typical DM
velocity dispersion is low.  If significant DM cores are found in these
galaxies, their existence would give support to models with a variable
SIDM cross section that is higher (20 cm$^2$/g, see eq.2) at small halo
masses and then declines rapidly at the scale of groups and galaxy
clusters as constraints at scale support small cross sections $\sigma$
$<$ 1 cm$^2$/g.  We plan to further explore the relative effect of
stronger SIDM interactions on satellites compared to field halos in
future work.

\begin{table}

\centering
\begin{tabular}{|p{0.2cm} p{1.6cm} p{0.6cm} p{1.6cm} p{1.6cm}|}
\hline
Run        & ~Halo Mass & V$_{max}$ &Stellar Mass  & ~~HI mass  \\
 ID        &   CDM/SIDM   &  \kms  &  CDM/SIDM      & ~CDM/SIDM  \\
\hline
h516 &   ~~~~4.1~$\times$10$^{10}$ & 58.4 &6.2/4.9~$\times$10$^7$  & ~~3.3/2.0~$\times$10$^8$\Msun \\
h516b &   ~~~~1.1~$\times$10$^{10}$ & 33.0 &9.8/14.7~$\times$10$^6$  & ~~1.6/0.94~$\times$10$^8$\Msun \\
h2003 &  ~~~~1.1~$\times$10$^{10}$  & 30.5 &3.1/3.4~$\times$10$^6$ & ~5.0/16.6~$\times$10$^6$\Msun\\

\hline
\end{tabular}
\caption{{\sc The  total halo mass, stellar and HI masses
 and V$_{max}$ (in \kms)} for  some representative halos in the CDM runs with SF and their  SIDM counterparts.
    z$=$0 stellar masses were  measured within 2.5kpc. h516b is the second most massive halo in the h516 run.}
\end{table}

\begin{figure}
\hspace{-0.5cm}\includegraphics[width=0.5\textwidth]{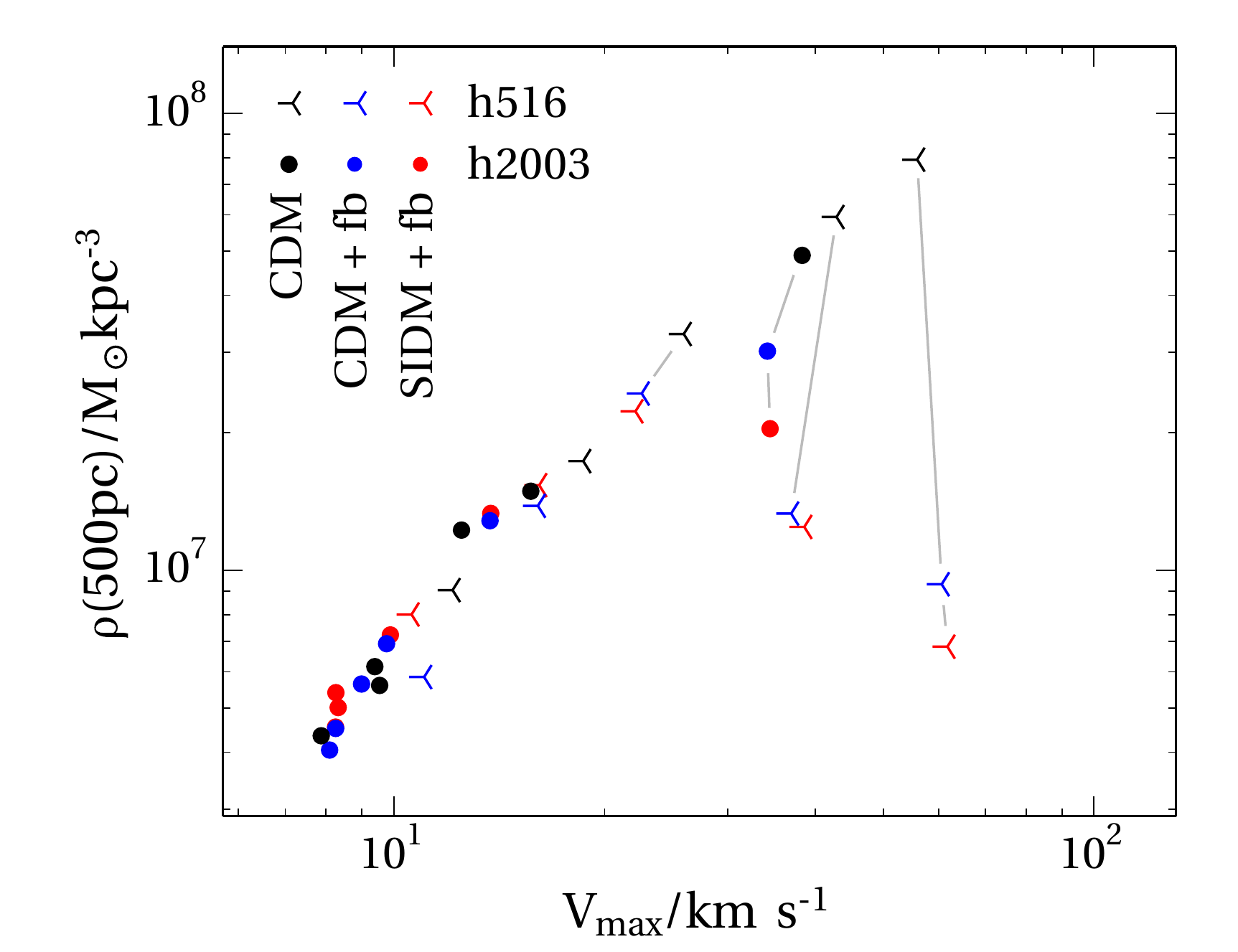}
\caption{{\sc Effects of Baryons Physics:} The central density of
  CDM-only halos (blue) vs their  SIDM (red) counterparts run
  with baryon physics and  bursty feedback. Lines connect the
  DM-only runs of h516 and h2003 to their counterparts with baryons.
  For our choice of the SIDM interaction cross section and for halos
  with V$_{max}$ $>$ 50 \kms (corresponding to stellar masses larger
  than 10$^8$ \Msun) the central mass of CDM and SIDM halos is
  similar, but significantly lower than the predictions of CDM-only
  runs. Due to low DM interaction rates and lack of bursty outflows, at
  lower halo masses both cosmologies give similar predictions: cuspy
  central DM profiles. Larger SIDM cross sections would be able to
  differentiate the various models in halos with V$_{max}$ $<$ 20\kms
  as  DM cores would then form in galaxies were baryon physics
  do not play a major role.}
\label{figbar}
\end{figure}

\begin{figure}
\vspace{-4em}
\hspace{-0.6cm}\includegraphics[width=0.525\textwidth]{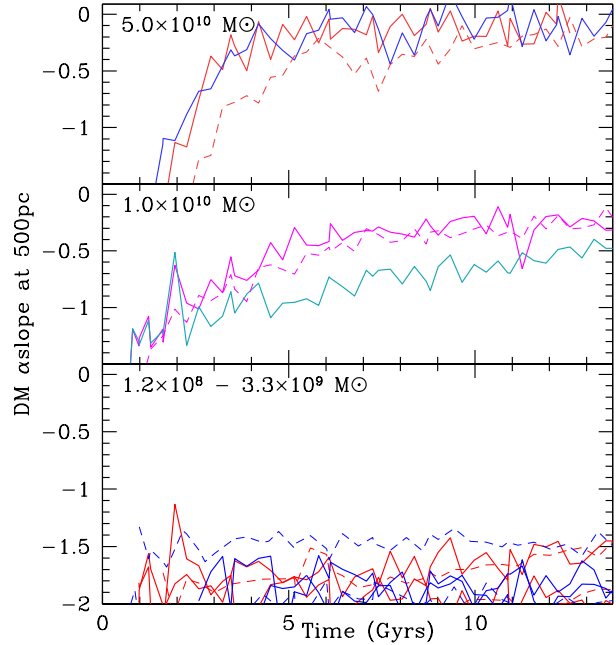}
\vspace{-9em}
\caption{{\sc The slope $\alpha$ of the DM density profile in SIDM and CDM over 
 different mass ranges:} blue and cyan:
 CDM, red and magenta: SIDM, solid lines: baryon+DM runs, dashed lines: DM-only runs. Top: halo h516. The DM
  slope evolves rapidly and in a similar way in both CDM and SIDM, as
  dynamical heating is very efficient at this scale.  Middle: halo h2003.
  (blue: CDM, red: SIDM). Bottom: a collection of field halos with
  total mass $<$ 10$^9$ \Msun ~from our simulations. SF efficiency is too low and 
the SIDM interaction rate is too low to  create cores.}
 \label{fig6}
\end{figure}

\subsection{Simulations with SF: SIDM similar to CDM when Baryon
  Physics are relevant. }
We focused our hydrodynamical simulations on the largest galaxies
formed in the filamentary regions `h516' and `h2003'.  These are two
well studied fields: h2003 is the same halo simulated in \cite{GWDM}
while h516 is the main halo of \cite{G10} and of the ``7 dwarfs''
galaxies sample studied in \citet{shen14} and \citet{madau14a}.  The
SF parameters in our study were identical for all CDM and SIDM
simulations. They also correspond to the fiducial ``g5'' runs in
\cite{GWDM}, where we explored the effects of different feedback and SF
recipes in the context of comparing the formation of dwarfs in CDM vs
Warm scenarios. Here we emphasize that our SF implementation creates
repeated starburts and gas outflows with significant loading factors
(gas mass ejected from the center divided by star formation rate).
Multi-wavelength evidence for outflows, analysis of the stellar
populations in the SDSS dwarfs and realistic CMDs \citep{GWDM} give
strong support to our implementation of SF.  This approach differs
from the SIDM study of \cite{vogelsberger14} where less bursty
feedback still removes gas from galaxy centers, but does not
create substantial DM cores.  The baryonic content of the galaxies in
our study are summarized in Table 2.

Figure \ref{figprofile} (left panel) compares the density profiles of the DM
component of galaxies h516 and h2003. In DM-only simulations halos
with masses $>$ 10$^{10}$ \Msun ~have cuspy profiles (halo h2003: cyan
dashed, halo h516: blue dotted).  Once baryon physics and outflows
are introduced, flatter DM profiles are created in both SIDM and CDM
cosmologies. The blue dashed (CDM-only) vs blue (CDM+SF) lines and the
red (SIDM+SF) show results for h516. The cyan dashed (CDM-only) vs
cyan solid (CDM+SF) lines and the magenta (SIDM+SF) lines show
results for h2003, the smaller halo.  In both CDM and SIDM models the
central cuspy profiles have been significantly flattened inside 1~kpc.
h516, the most massive halo studied with the inclusion of SF, is the
one where the central DM density decreases the most.  This result is
consistent with previous findings \citep{G12,dicintio14}, showing that
the efficiency of core formation peaks in halos with V$_{max}$ $\sim$
50 \kms.

\begin{figure*}
\hspace{-0.25cm}\includegraphics[width=0.95\textwidth]{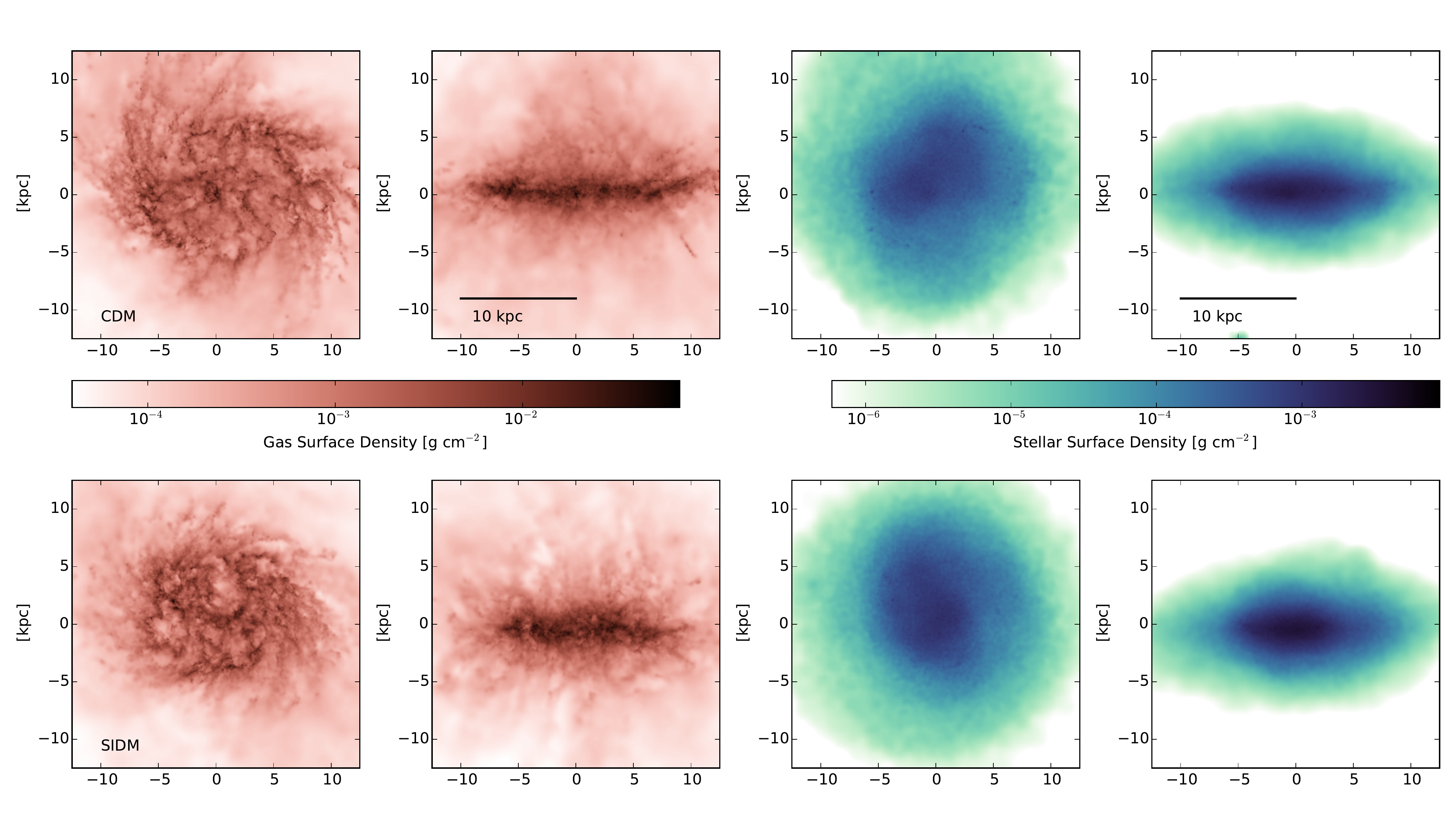}
\caption{{\sc The projected color density map of the baryon
    distribution in galaxy h516.} Top: CDM. Bottom: SIDM. Left panels:
  projected gas density at z$=$0 seen edge-on and face-on. Right panels:
  Stellar projected density at z$=$0. The stellar disks are relatively
  thicker than in earlier simulations \citep{G10}, an effect of the
  introduction of `early feedback' from young stars. We verified that
  in both models the stellar distribution is exponential, with a similar
  disc scale length and lack of a central dense spheroid.}
\label{baryon map}
\end{figure*}

Figure \ref{figprofile} (right panel) compares instead the velocity
dispersion profiles of the DM component in halos h516 and h2003.  As
for the density profiles, DM-only runs of different cosmologies have
different local DM velocity distribution profiles: CDM halos show a
decreasing dispersion closer the halo center while the SIDM halos show
a rather flat profile. This difference, a result of the energy
transfer to the center of the halo due to collisional processes, had
been considered a strong signature of SIDM.  However this difference
is erased by the introduction of bursty feedback, that creates
significant DM cores in the CDM halos. `Dynamical heating' also causes
the velocity dispersion profiles of all halos  to flatten
out.

Figure \ref{figmass} shows the stellar mass/halo mass ratio of
  the halos we simulated with the inclusion of baryon physics, showing
  that they follow the stellar mass/halo mass relation inferred using
  local data and the abundance matching technique
  \citep{brook14,gk14}. This is particularly relevant as producing the
  correct amount stars is necessary to estimate the minimum mass at
  which baryonic processes can originate cores.  After this paper was
  submitted \citep{onorbe15} published results from a simulated CDM
  halo of similar total mass, which also formed abut 10$^6$ \Msun in
  stars, leading to a cored DM profile.

Figure \ref{figbar} shows the DM central densities as a function
of halo peak velocity once SF is included. The comparison with
Figure \ref{fig1} is striking: where at V$_{max}$ $>$ 30 \kms CDM was
clearly differentiated from SIDM, now the central density of CDM and
SIDM halos is almost identical over the whole 10 to 50 \kms
range. This result clearly illustrates how predictions form DM-only
runs can be completely superseded by the addition of the complex
baryon -- DM interactions.

In Figure \ref{fig6} we investigate how the slope of the DM profile (measured
at 500~pc) evolves with time in a sample of field halos with a range
of masses.  As the mechanisms of core formation differ (potential
fluctuations linked to SF vs SIDM collisions) the time evolution of
the DM slope may also be significantly different. We have found that the
evolution of the systems can be roughly divided in three stellar mass
ranges.  Present day systems with stellar masses larger than 10$^8$
\Msun ~(V$_{max}$ $>$ 50\kms as in classic field dwarfs) had cored DM
profiles since redshift $>$ 4, with no significant difference between
SIDM and CDM (top panel). In the SIDM-only run, the DM core forms more
slowly than when the effect of baryons are included. The effect of
baryon feedback on the DM profile is clearly detectable as soon as about 10$^6$ \Msun ~of stars have been
created \citep[see also][]{GWDM, penarrubia12}.  In galaxies with
present day stellar masses below 10$^{7-8}$ \Msun, the effect of
`dynamical heating' induced by feedback is progressively reduced 
\citep{G12,GWDM,dicintio14}. At the radial scale of 500~pc the CDM
and SIDM halos of h2003 show DM profiles that start cuspy and then
slowly develop a flatter profile. In both SIDM and CDM the process is
more gradual compared to more massive systems, due to a number of factors
(star formation rates are lower and $\tau_{\mathrm{SI}}$ is longer). However, the
inclusion of baryonic processes again makes the galaxy evolve
similarly in CDM vs SIDM. By redshift z$<$1 the difference in DM
slopes is not significant enough to discriminate between SIDM and CDM.
Halo h2003 was also recently run in a WDM cosmology
\citep{GWDM}. Similar to the CDM case the WDM cusp turned into a core
over time, due to bursty feedback. By z$=$0 the central DM distribution of
halo h2003 is similar in CDM, WDM \citep{GWDM} and SIDM (this work).

\begin{figure*}
\centering{
\raisebox{-28ex}{\hspace{-0.0cm}\includegraphics[width=0.525\textwidth]{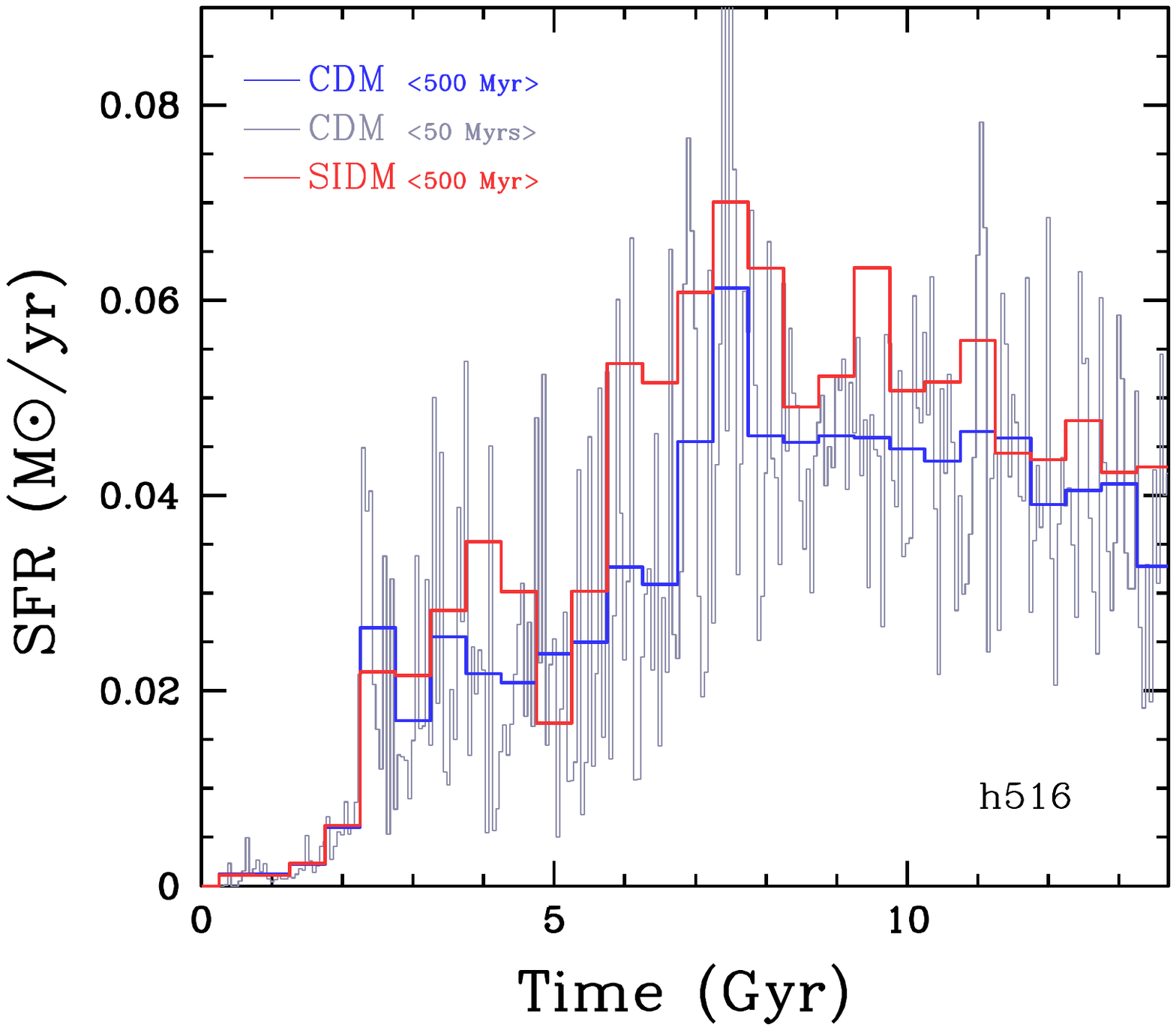}}
\raisebox{2ex}{\hspace{-0.8cm}\includegraphics[width=0.475\textwidth]{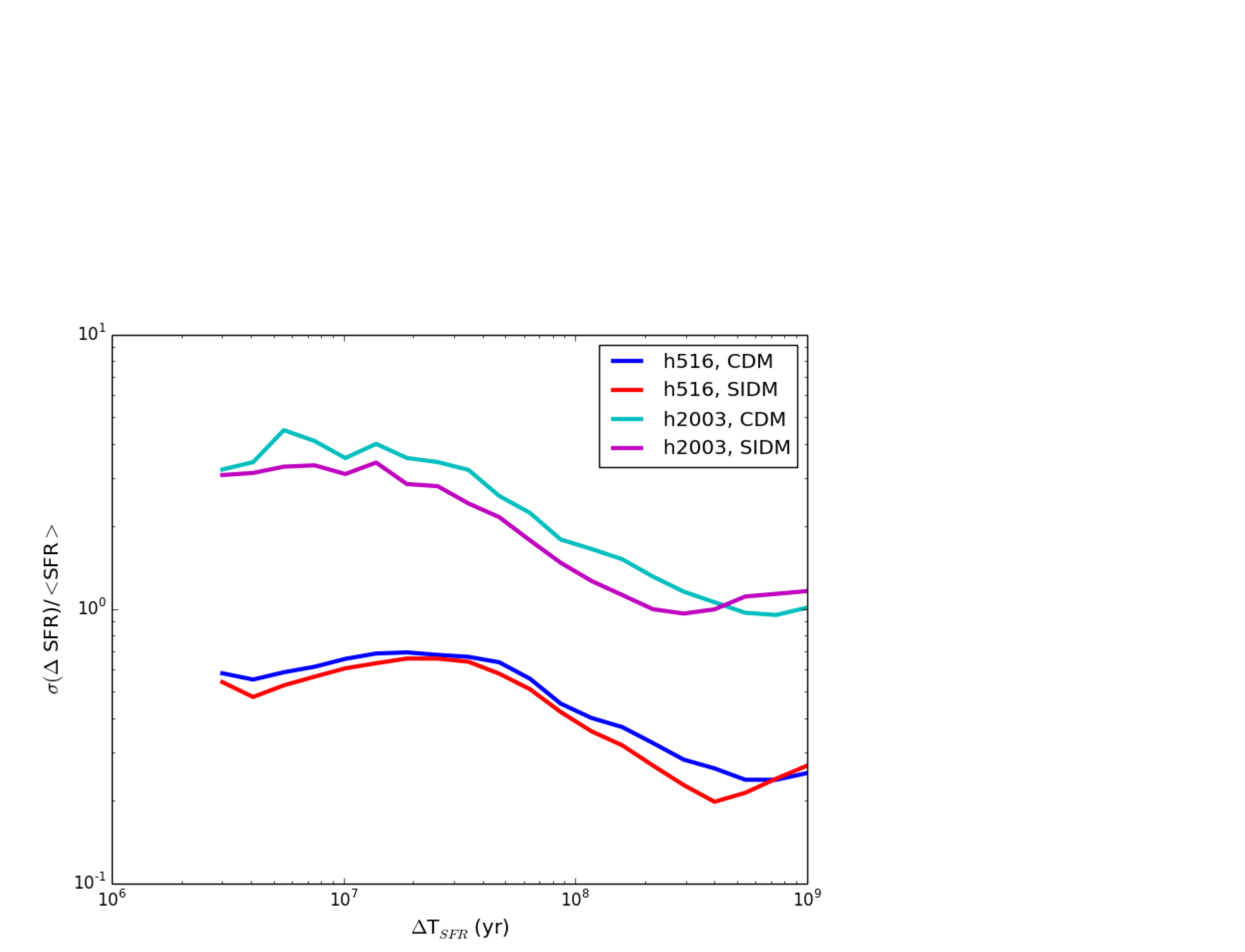}}
}
\vspace{-12em}
\caption{{\sc The SFR as a function of cosmic time}.  Left panel: halo
  h516 in CDM and SIDM cosmologies.  Both galaxies have extended, gas
  rich disks and negligible bulges. Right panel: the burstiness of
  models h516 and h2003 in CDM and SIDM, measured at different
  timescales (see text for a definition). SF does not differ
  significantly in CDM vs SIDM, and appears to be mostly regulated by
  feedback and accretion rates.}
\label{fig8}
\end{figure*}

At even smaller masses, the evolution of the DM slope $\alpha$ with the
inclusion of baryonic processes (continuous lines) confirms the
results from the DM-only runs (dashed lines). SF at such small scales
is strongly inhibited by the cosmic UV field. Similarly, at such small
masses the SIDM interaction timescale becomes long compared to the
Hubble time and the survival time of the halo. Hence the central DM
slope measured at 500pc is not significantly different in SIDM vs CDM
halos, as predicted by the empirical calculations in the
previous section. The failure of SIDM to form substantial DM cores in very faint
dwarfs with host halos with mass $<$ 10$^{10}$ \Msun ~had not been
explored in previous numerical works, where high resolution
simulations mostly focused on halos more massive than 10$^{10}$ \Msun.

{\it Overall, the inclusion of SF and baryonic processes erase the
  differences seen in the DM-only runs. At the scale of dwarf galaxies
  with V$_{peak}$ $\sim$ 30-50 \kms. energy transfer to the DM makes
  the DM distribution at the center of galaxies flatter than an NFW
  profile and almost identical in CDM vs SIDM }.  As a result current
constraints on the SIDM cross section from dwarf sized field come out
considerably weakened and will have to be reconsidered. Moreover, we
predict (see Figure \ref{fig1}) that a SIDM cross section $>$ 10
cm$^2$/g would necessary to create DM cores in smaller halos with
V$_{max}$ $<$ 10 \kms. As the shapes of galaxy clusters require a SIDM
cross section smaller than 1 cm$^2$/g \citep{peter2013}.  If cores
exist in dwarf galaxies, these results argue in favor of a variable
SIDM cross section (see also \cite{vogelsberger12}).

\subsection{Star Formation Histories and Assembly of the baryonic component}

In this section we analyze the $z = 0$ baryonic distribution of the
galaxies that form stars in our simulations, while focusing on the two
main halos: h2003 and h516.  We verified that in both CDM and SIDM
cosmologies the stellar mass of galaxy h516 follows an exponential
profile with no spheroidal component.  The projected gas and stellar
densities of galaxy h516 are shown in Figure \ref{baryon map}, showing
a similar distribution in CDM vs SIDM.  Both stellar discs are thick
and dynamically hot (Figure \ref{baryon map}) and more extended than
the CDM version of h516 described in \cite{christensen14}, an effect
of the introduction of early feedback \citep{trujillo13,roskar14}.
Confirming the visual impression from Figure \ref{baryon map}, we
verified that the radial distribution of stars at z=0 does not change
significantly in the CDM galaxies vs their SIDM counterparts, both
following a trend for larger systems to have more extended stellar
systems and roughly consistent with observational data. The half light
radius for h516b (Table 1) is for example, 2.0 kpc in SIDM, vs 1.8 kpc
in CDM.  This is a different outcome compared to the results of a less
bursty SF scenario as the one studied in \cite{vogelsberger14}, which
followed galaxies in a similar mass range to ours. In their
simulations the SIDM distribution is less concentrated compared to the
CDM one, leading ``to dwarf galaxies with larger stellar cores and
smaller stellar central densities.'' In our simulations the dynamical
coupling of baryons to DM, driven by fast and repeated outflows,
shapes the stellar distribution and leads to a similar baryon
distribution and content.

\begin{figure*}
\centering
\includegraphics[width=0.47\textwidth]{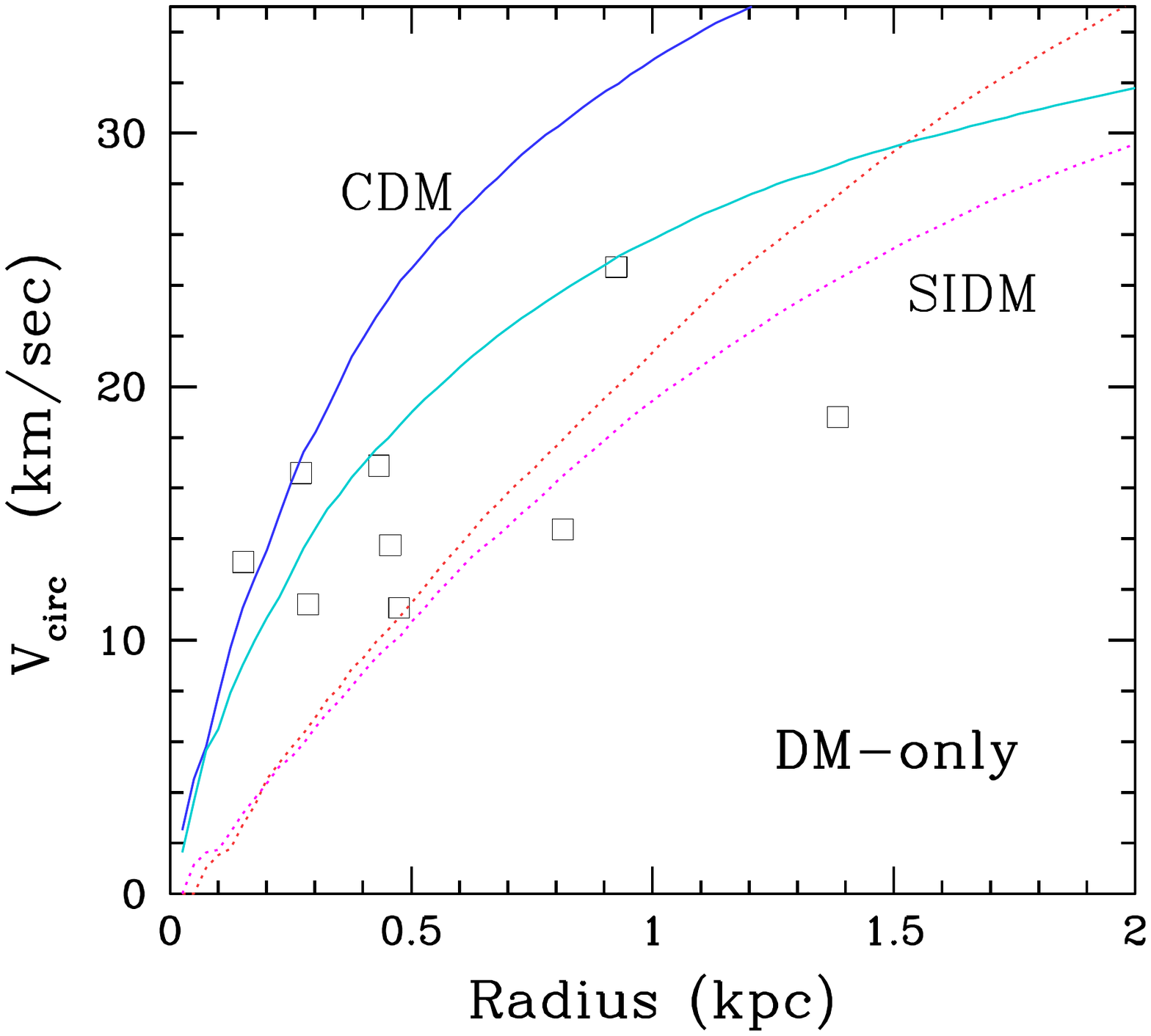}
\includegraphics[width=0.47\textwidth]{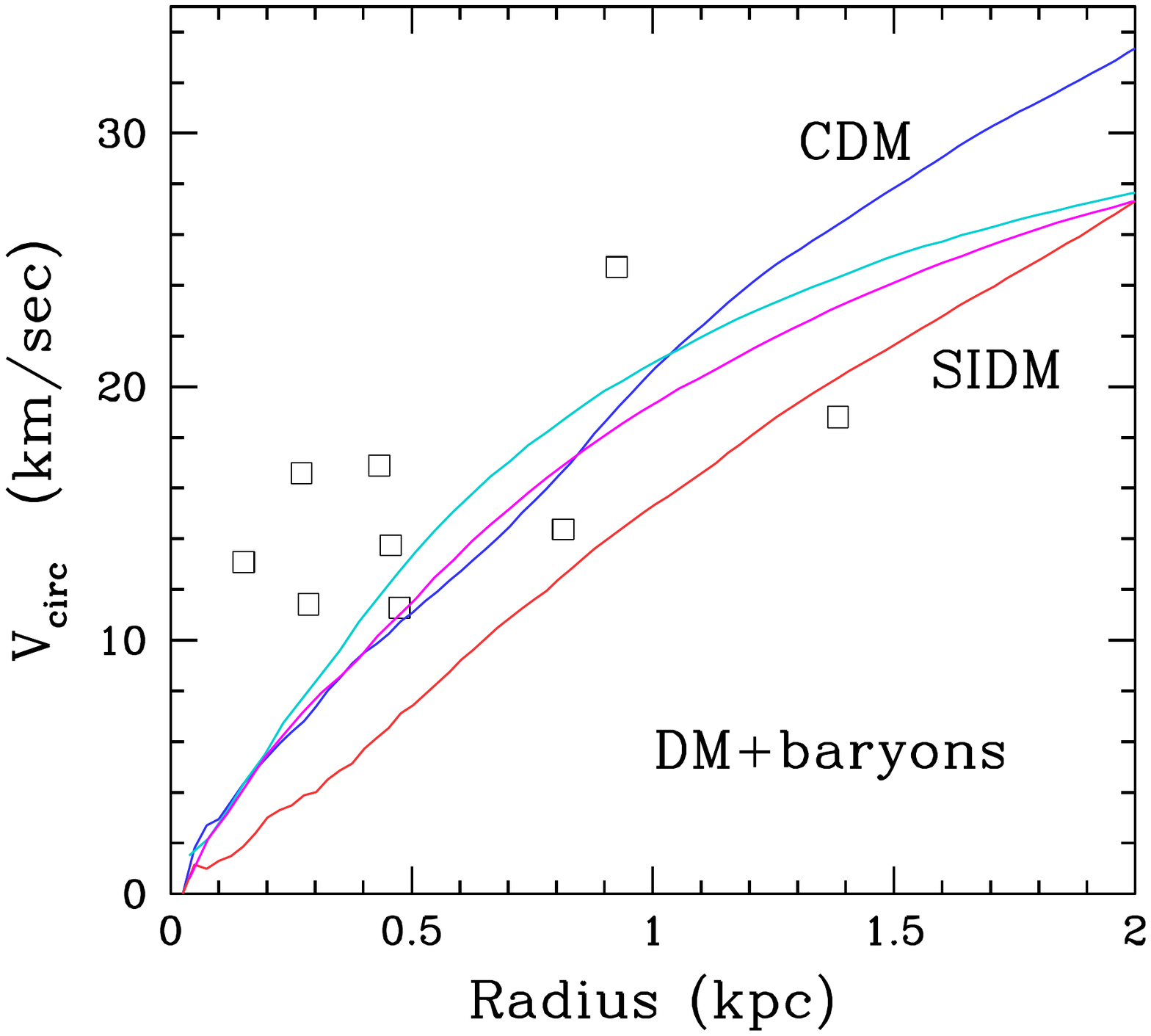}
\vspace{-8em}
\caption{{\sc The rotation curves V$_c$} of halos h516 and h2003 in
  CDM (blue and cyan) and SIDM (red and magenta) compared with the
  observational constraints from a sample of Local Group field dwarfs
  \citep{wolf10,Mcconnachie12,weisz14}. The introduction of feedback
  processes lowers V$_c$ out to at least 2~kpc, where mass
  decomposition based on HI kinematics are crucial. (V$_c$ is defined as  = $(GM/r)^{1/2}$). Softening
is 64pc for h2003 and 86pc for h516.}
\label{fig9}
\end{figure*}

Furthermore, with the adopted SF prescription, the star formation
histories (SFH) of the galaxies formed in the CDM model look very
similar to the SIDM counterparts. This is most likely due to two main
reasons: 1) the assembly rate of both DM and baryons, being driven by
the large scale structure remains unchanged in SIDM and 2) there is no
difference in the central DM distribution between the two models as it
is responds quickly to the effects of feedback.  As an example, Figure
\ref{fig8} shows the SFH of halo h516 in the CDM and SIDM models,
which show a very similar time evolution, peaking 5 Gyrs after the Big
Bang.  Overall, SIDM and CDM halos have almost identical DM masses by
z$=$0 and the SF efficiency is similar, within 20\% in the two
models. This result differs from the CDM vs WDM comparison in
\cite{GWDM}, where the WDM version of h2003 formed only half the stars
due to the delayed assembly of the halo \citep[see also][]{colin15}.
However, changes on the timescale and intensity of the individual SF
events could be driven by subtle effects and still differentiate
between CDM and SIDM, for example a shallower density profile could
drive more gas instabilities that lead to radial inflows and increase
the burstiness of SF, or, on the other hand, lower DM densities could
slow down the collapse of gas and reduce SF in the SIDM scenario. We
quantified the burstiness of SF in SIDM vs CDM by measuring the
dispersion in star formation rates (SFR) measured in $\Delta$T$_{SFR}$
intervals, where $\Delta$T$_{SFR}$ was varied from 10$^6$ to 10$^9$
years \citep[for a similar approach but different definition of
burstiness, see][]{hopkins14}. Measuring $\Delta$T$_{SFR}$ over
hundreds of randomly sampled intervals, bursty SF at a given time
scale shows as a large dispersion if the SF rate at a time
$T+\Delta$T$_{SFR}$ differs substantially from the SFR at a previous
time $T$. The right panel of Figure \ref{fig8} shows that the
burstiness measured over different timescales was similar for both
halo h516 and h2003 in both models.  Both h516 and h2003 show a SF
rate dominated by small time scale fluctuations, with a similar
dependence on time scale and halo mass, with the less massive halos
having a burstier SF at all time scales \citep[a result seen also in
simulations with different feedback models, e.g.,][]{hopkins14}.
Observations of nearby stellar populations and comparisons with
simulations \citep{tolstoy09,kauffmann14,GWDM,weisz14} strongly support
a bursty build up of the stellar content of dwarf galaxies. We
  verified that our set of simulations qualitatively reproduce the
  fraction of stars formed in bursts as estimated in
  \cite{kauffmann14}): 20-50\% with a larger fraction formed in the
  less massive system. Bursts were defined as periods with SF $>
  2-4\times <SFH>$ and duration of 10$^7$-10$^8$ years. A burstier
  implementation of our SF model (as one that neglected HI self
  shielding) would be ruled out by excessive structure in the stellar
  CMD, as discussed in \cite{GWDM}.

 {\it Overall, Within our small sample
  of galaxies, we do not find evidence of large differences in the
  baryonic content and distribution in galaxies formed in a CDM vs
  SIDM model.}

Finally, Figure \ref{fig9} show the rotation curves V$_c$ (defined as
$\sqrt{GM(<r)/r}$) of our galaxies h516 and h2003 compared to
estimates from a sample of local dwarfs \citep[empty squares,
from][]{wolf10, weisz14}. The left panel shows the well known result
that halos formed in CDM-only simulations have higher central mass
densities that most observed dwarfs (although significant
uncertainties remain on the observational estimates). At the same time,
SIDM-only runs with a range of cross sections \citep{elbert14} show a
better agreement with observational estimates as the central DM
densities are lower.  However, the right panel of Figure \ref{baryon map}
shows that once baryon physics are introduced, the central mass
distribution (DM$+$baryons) of CDM dwarfs matches the observational
data, equally well, without requiring SIDM. Recent works have compared
dwarf galaxies in the Local Group \citep{brook14,sawala15} and MW
satellites \cite{zolotov12,arraki14} with simulations. They have shown
how the introduction of feedback (and tidal processes for satellites)
generates a realistic M$_{star}$ -- M$_{halo}$ relation, as galaxies
at a fixed M$_{star}$ are hosted by more massive halos than when a
universal NFW profile is assumed in the analysis \footnote{in a halo
  model with a central mass density lower than NFW the V$_{circ}$
  measured at $\sim$ 1~kpc translates to the same V$_{peak}$ and hence
  halo mass of an NFW halo.  As a consequence, an observed V$_{circ}$
  translates to higher halo host mass than when a more concentrated
  NFW halo is assumed.}. This result shows how the perceived
discrepancy between the observed number density of field galaxies as a
function of V$_{max}$ and that of the underlying halo population is
naturally accounted for in a CDM context where feedback is introduced
and the central total density is lowered. In good agreement with
results from our simulations, the revised abundance matching from
\citet{brook14} predict that field galaxies are hosted in halos with
mass $>$ 5 $\times$ 10$^9$ \Msun, with most halos below that threshold
being devoid of stars. Our results provide direct evidence that
SIDM-only simulations mimic the effects of feedback, but also that
{\it when bursty feedback is introduced, SIDM galaxies are essentially
  indistinguishable from their CDM counterparts.}  Both SIDM and CDM
models create galaxies with mass distributions and observable
properties in broad agreement with the observed ones.

\section{Conclusions}

We studied a set of cosmological simulations of field dwarf galaxies
in CDM and SIDM. The SIDM simulations include for the first time a
description of bursty SF and feedback that creates potential
fluctuations able to lower the central DM density of halos where at
least 10$^6$ \Msun ~of stars have formed.  The force resolution of our
simulations (64-86 pc with a spline kernel softening) allows us to
resolve with many thousands of particles the central regions of halos
(and millions to the virial radius) down to halo masses where the
impact of star formation will not play a major role and where the
effect of SIDM alone dominates. A relatively large value (2
  cm$^2$/g for the SIDM cross section was chosen to maximize its
  difference from the CDM model. We have also run a set of DM-only
simulations that allowed us to study the evolution of a uniform
resolution set of DM halos down to masses of only few times 10$^8$
$M_{\odot}$ (equivalent to V$_{max}$ $<$ 20 \kms) both in the field
and in the dense environment of a more massive host.  Our results can
be summarized by three main findings:

\begin{itemize}

\item Once SF and resulting feedback is introduced, the central DM
  mass distribution and velocity dispersion becomes similar for CDM
  and SIDM galaxies with stellar masses in the 10$^6$--10$^8$ \Msun
  ~range and circular velocity 25-50 km/sec (Fig.3).  At the scale of
  0.5--2 kpc the total matter content is in good agreement with
  observational estimates of Local Group dwarfs.  This result differs
  starkly from the predictions of DM--only simulations, or simulations
  where energy feedback does not lead to core formation
  \citep{bk11a,gk14}.

\item Analytical expectations and simulations show that with a fixed
  SIDM cross section of 2 cm$^2$/g the DM central density and internal
  velocities of halos with mass below 10$^9$ \Msun ~are too low to
  have a significant number of DM--DM interactions. Such SIDM halos
  remain cuspy when observed at a scale of 500pc, and show an enclosed
  DM mass content lower by about only a factor 2 compared to their CDM
  counterparts (Fig.1 \& 2).  This result poses an interesting {\it
    lower limit} to the SIDM cross section in halos associated with
  galaxies with stellar masses below 10$^6$ \Msun.  If kpc sized DM
  cores are found in these galaxies, their existence would give
  support to models with a variable SIDM cross section that is high (
  $\sim$ 10-20 cm$^2$/g, see eq.2) at small halo masses and then
  declines rapidly at the scale of galaxies and clusters.  Large
    cross sections at dwarf scales are also suggested by DM-only
    studies \citep{elbert14}.  Such large cross sections may have a
    detectable effect on the SFHs of dwarf sized systems, but could
    also make galaxy satellites easier to disrupt.  Our simulations
  also highlight the relative environmental effects of the enhanced
  interaction rate at the center of very small satellites compared to
  their field counterparts, due to rapid orbital velocities in the
  dense halos of a massive host. This relative difference offers the
  potential of using faint galaxies to constrain the cross section of
  SIDM as a function of the DM velocity.
\item Once regulated by feedback, the SFHs (Fig.7), stellar and gas
  content and spatial distribution do not differ substantially in CDM
  vs SIDM model, both forming gas rich galaxies with bulgeless disks
  which resemble real ones.  

\end{itemize}

These results lead to potentially important implications for SIDM
models. As baryons lead to a drastic change in the mass distribution
of central regions of typical dwarf galaxies with V$_c$ $\sim$ 30-60
km/sec, many of the current bounds on the SIDM cross section at
different mass scales will need to be re-examined. After the
introduction of realistic descriptions of SF and feedback. Both CDM
and SIDM form dwarf galaxies with DM distribution and observable
properties in broad agreement with observational data.  As a result,
while certainly not excluded, SIDM is not necessary to solve the
problem of the excess of DM at the center of simulated dwarf galaxies
with V$_{max}$ $>$ 30 \kms.  Furthermore, because the effect of SIDM
becomes negligible at very small galaxy/halo masses even with a
relatively large constant velocity cross section (2 cm$^2$/g), SIDM
with a velocity dependent cross section will become necessary if even
very faint dwarf galaxies are observed to have dark matter cores.
This work highlights how simulations that a) include realistic SF and
feedback processes and b) study a large sample of faint dwarfs are
necessary to take advantage of astrophysical constraints on DM models.

\medskip
\section*{Acknowledgments}
ABF was funded by Washington NASA Space Grant Consortium, NASA Grant
NNX10AK64H. FG and TQ were funded by NSF grant AST-0908499.  FG
acknowledges support from NSF grant AST-0607819 and NASA ATP
NNX08AG84G. AP is supported by a Royal Society University Research
Fellowship. Some results were obtained using the analysis software
pynbody, \citep{pynbody}. ChaNGa was developed with support from
National Science Foundation ITR grant PHY-0205413 to the University of
Washington, and NSF ITR grant NSF-0205611 to the University of
Illinois. We thank Haibo Yu, Sean Tulin, Matt Walker and Lucio Mayer
for stimulating discussions and a careful reading of the
manuscript. We thank the referee for constructive comments.

\bibliography{bibref}

\bibliographystyle{mn2e}

\section{Appendix: The SIDM implementation}

\begin{figure}
\centering
\includegraphics[width=0.425\textwidth]{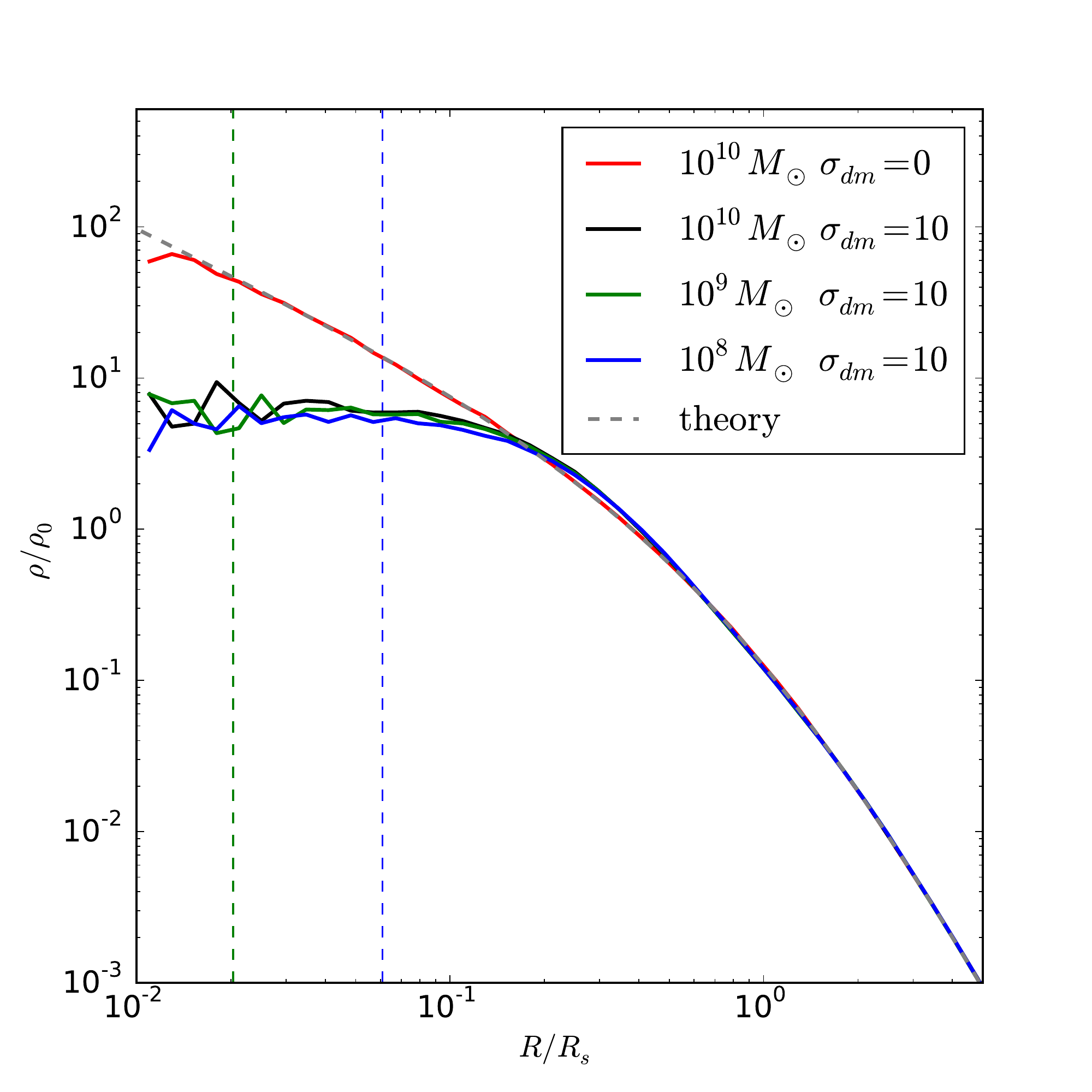}
\caption{The density in scaled units of $R / R_{s}$ for Hernquist halos of 10$^{10}$ (black), 10$^{9}$ (green), and 10$^{8}$ \Msun
  (blue) which have scale radii of $1.099$, $.37$, and $.12$
Kpc respectively based on the mass concentration ratio relation from Neto et
al 2007. Each halo has been evolved for about 25 scaled dynamical times which corresponds to 91, 44, and 20 million years respectively. The
halos demonstrate self similar evolution in these scaled distance and time
units. The density $\rho$ has been scaled by the central density.
The vertical dashed lines indicate the softening length of 7.5 pc that
is scaled for each halo (the dashed line for the $10^{10}$ \Msun halo is
off the plot further to the left). The dashed grey dashed line is the Hernquist density profile.
}
\label{figA1}
\end{figure}

\begin{figure}
\hspace{-0.7cm}\includegraphics[width=0.55\textwidth]{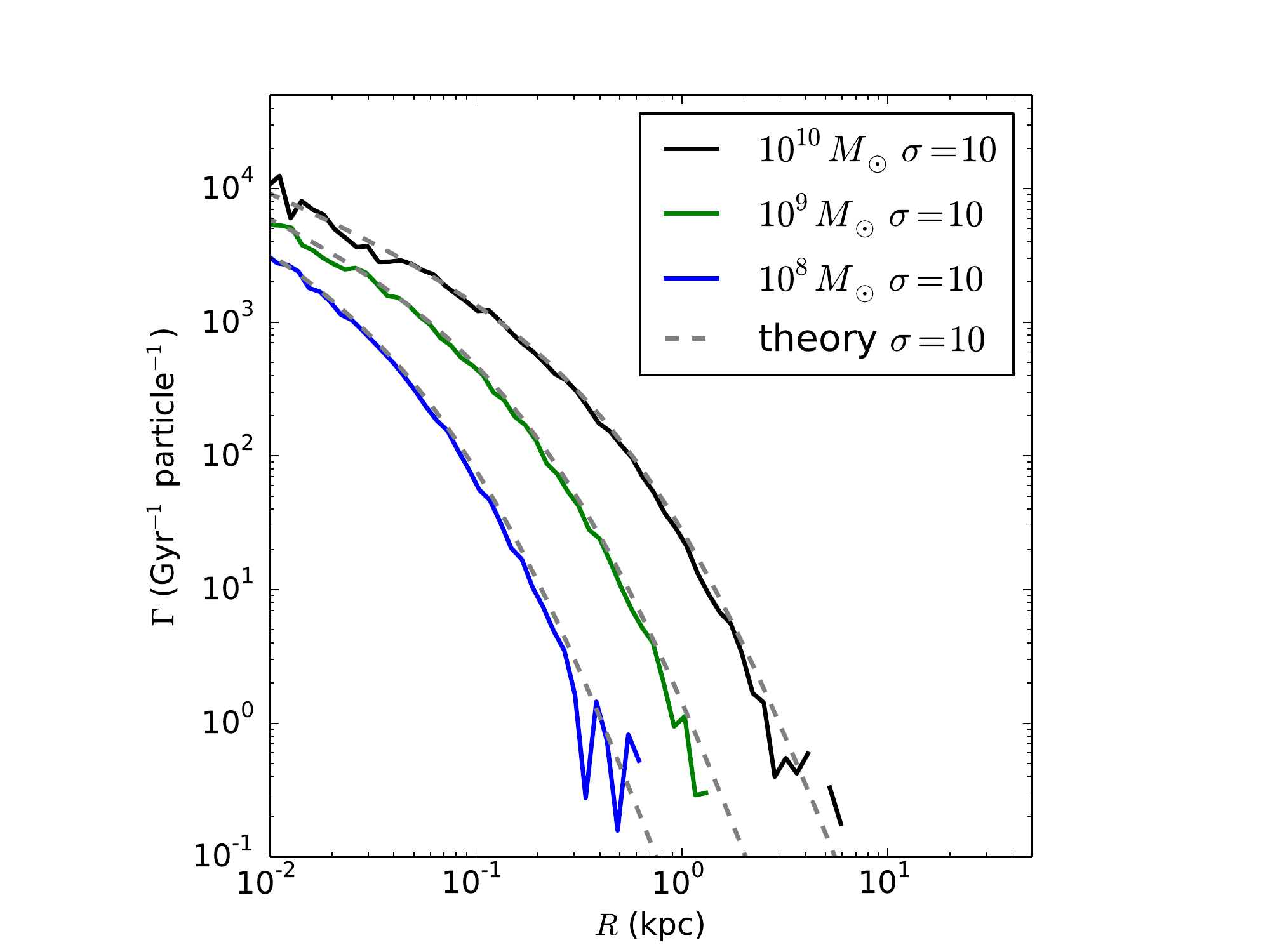}
\caption{ Snapshots of a Hernquist profile SIDM simulation showing the number of interactions that occur per particle per giga-year
  for the same halos as above. The cross section is 10 cm$^{2}$ g$^{-1}$. In dashed grey is
  the analytical prediction for each halo respectively. The simulations were run for a period much shorter than the dynamical time of the halo. Each Hernquist profile has a different
  scaled structure as described in the previous figure.
  Each halo was described with about 2 million particles.
}
\label{figA2}
\end{figure}

We verified our implementation of SIDM by comparing analytical
predictions  of the number of dark matter interactions
to simulations of dark matter halos with a Hernquist density
profile \citep{hernquist90} (Fig. \ref{figA1}).  We then demonstrate
the formation of a flat core and the signature flat SIDM velocity
dispersion profile, starting from an NFW profile with a cuspy density
profile and a raising velocity dispersion \citep{navarro96}.  Our
implementation code closely replicates published results of the
predicted number of SIDM collisions as a function of halo mass and
local density and the evolution of the central density profile \citep{vogelsberger12,rocha13}
(Fig.\ref{figA1} and Fig.\ref{figA2}).

The number of dark matter
interactions per unit time is an important cosmological value for SIDM
theories and a useful diagnostic for  numerical implementations of SIDM interactions
The probability of an interaction between any two
particles is given in eq.4. The total
number of interactions, $\Gamma$, that occur per unit time is an
integral of the velocity weighted cross section in the volume $V$.
\begin{eqnarray}
\Gamma= \int \frac{\rho^{2}(\textbf{x})}{2 m_{\chi}} \langle \sigma _{dm} v_{r} \rangle (\textbf{x}) dV
\end{eqnarray}
Where $\rho$ is the local DM density, $m_{\chi}$ is the mass of the dark matter particles in the
simulation, $\sigma_{dm}$ is the dark matter cross section per gram, and $\langle
\sigma_{dm} v_{r} \rangle(\textbf{x})$ is the local thermal average of the cross
section weighted by the particle's relative velocity as a function of
the position $\textbf{x}$. The local thermal average of the cross section is
calculated by taking the first moment of $\sigma_{dm} v_{r}$ over the combined relative velocity distribution function $f (v_{r})$.

\begin{eqnarray}
\langle \sigma v_{r} \rangle (\textbf{x})= \int^{\infty}_0 (\sigma_{dm} v_{r}) (\textbf{x})f (v_{r}) dv_{r}
\end{eqnarray}
In the most general case the
distribution function can be a function of position (as for a dark
matter halo) and the cross section can be a function of relative
velocity (for velocity dependent SIDM). In the simplest case for
single value initial velocity of $v_{0}$ and uniform density, the
distribution  function $f(v_{r})$ becomes a delta function at that
given velocity where $f(v_{r})=$ $\frac{\delta(|v|-v_{0})}{4 \pi
  v_{0}^{2}}$ and the interaction rate simplifies to $\Gamma= \sqrt{2}/2 N
\rho \sigma_{dm} v$ where N is the total number of dark matter
particles in the simulation. A standard Maxwell-Boltzmann distribution has a constant velocity dispersion so that after averaging over all angles of collision we conclude $\langle v_{r} \rangle = \sqrt{2} \langle  v \rangle $ so that if $a$ is the standard deviation of the velocity vector in the distribution then the thermal average cross section is

\begin{eqnarray}
\langle  \sigma_{dm} v_{r} \rangle  (\textbf{x})&=& \sqrt{2} \sigma_{dm} \sqrt{\frac{8a^{2}}{ \pi}}
\end{eqnarray}
Thus in a constant density region with a Maxwell-Boltzmann velocity distribution the number of interactions per unit time is
\begin{eqnarray}
\Gamma= \frac{\sqrt{2} N \rho \sigma_{dm}}{2 } \sqrt{\frac{8a^{2} }{ \pi}}
\end{eqnarray}
We would like to have an analytic calculation for the number of interactions that would occur in a given dark matter halo, however in general the integrals of dark matter halos are not well behaved. Further, the dark matter interactions modify the density and velocity distribution of the halo by conducting energy (as a measure of velocity dispersion) from the outer regions of halo into the core. We approximate that after 25 scaled dynamical times $t_{0}^{-1}= \sigma_{dm} m_{\chi} R_{s} \sqrt{64 G \rho_{0}^3}$ (where $\rho_{0}=M/(2 \pi R_{s}^{3})$ is the central density) a core-collapse phase may start \citep{koda11}. One well behaved halo profile is the Hernquist profile which has a dark matter density profile similar to an NFW profile. In figure \ref{figA1} we show scaled Hernquist halos with masses of $10^{10}$, $10^{9}$, and $10^{8}$ $ M_{\odot}$ that have each been evolved for 25 scaled dynamical times. Under the assumption of an isotropic velocity distribution, the velocity dispersion  $\sigma_{v}$ of the Hernquist profile is known analytically from the literature and the average of $\langle  \sigma_{dm} v_{r} \rangle  (\textbf{x})$ is then

\begin{eqnarray}
\langle  \sigma_{dm} v_{r} \rangle  (\textbf{x})=\frac{1}{2 \sqrt{\pi} \sigma_{v}^{3}(\textbf{x})} \int^{\infty}_0 (\sigma_{dm} v) v^{2} \text{exp}[\frac{-v^{2}}{4 \sigma_{v}^{2}(\textbf{x})}] dv
\end{eqnarray}

The total number of interactions per unit time is then given by
equation 4 in the Appendix. However because the initial velocity
dispersion is that of a Hernquist profile in equilibrium, this
analytic prediction is an approximation which declines in accuracy
with halo evolution in time. In figure \ref{figA1} we show the
analytic prediction of the expected number of DM interactions compared
to a simulation of the halos that has been run for a brief period
(10$^5$ years), much smaller than the dynamical time of the halos and
before the density profile evolves significantly.

\subsection{Convergence of SIDM profiles in DM-only simulations}

\begin{figure}
\centering
\includegraphics[width=0.5\textwidth]{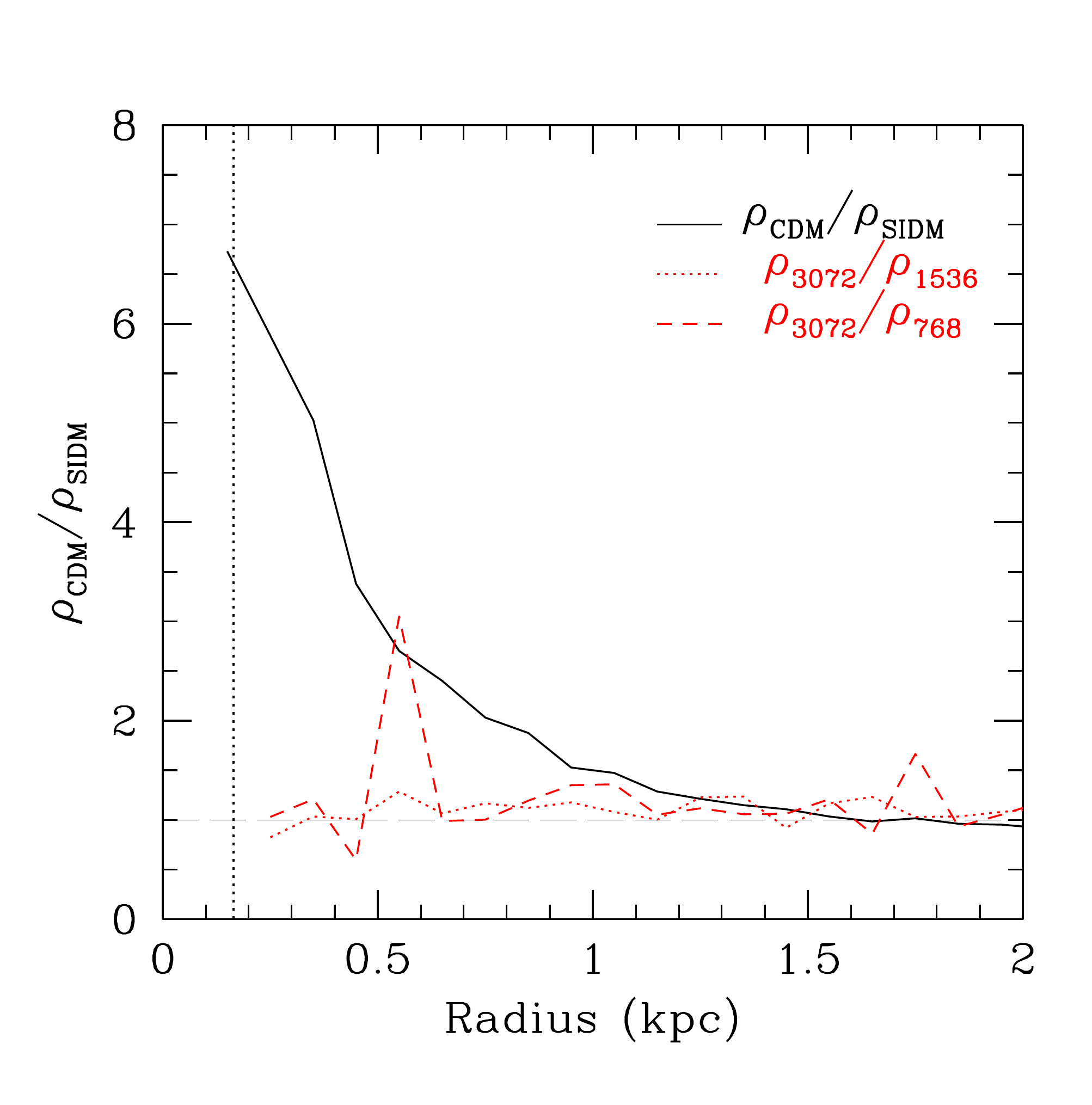}
\caption{The local density ratio as a function of radius between the
  high resolution SIDM version of halo h516 ($\rho_{SIDM}$) and its
  lower resolution counterparts ($\rho_{1536}$) and ($\rho_{768}$)
  Results in the low res versions converge at $\sim$ two softening
  lengths, as typical of  CDM-only simulations.  }
\label{resolution}
\end{figure}

In Figure \ref{resolution} we show a resolution test using the most
massive halo of simulation `h516' The halo (of final mass 4 $\times$
10$^{10}$ \Msun was simulated lowering the mass resolution by a factor
of 64 in mass and 4 in force resolution. The ratio of the spherically
averaged local density as a function of radius show that the SIDM
simulation converges at about 2 softening lengths, similar to the CDM
run (see also \cite{power03}) The `central' density decreases by a
factor of $\sim$ 6 between the CDM and SIDM case. These results agree
with previous works. To verify the convergence of the SIDM density
profiles at the previously poorly explored regime of M$_{vir}$ $<
10^9$ \Msun we also simulated the `40 Thieves' volume first with
particle mass 8000 \Msun and then again 2400 \Msun, and a force
resolution of 65pc in both cases. Combining these runs with the h148
volume we are able to cover four orders of magnitude in halo mass,
from small halos with peak velocities smaller than 10 \kms to massive
galaxies with a large system of satellites. We have simulated h148 at
lower resolution with hydrodynamics and SF, finding that it hosts a
large disc galaxy.

\bsp

\label{lastpage}

\end{document}

%% file: rgb.tex
  \definecolor{snow}{rgb}{1.000000,0.980392,0.980392}
  \definecolor{ghost white}{rgb}{0.972549,0.972549,1.000000}
  \definecolor{GhostWhite}{rgb}{0.972549,0.972549,1.000000}
  \definecolor{white smoke}{rgb}{0.960784,0.960784,0.960784}
  \definecolor{WhiteSmoke}{rgb}{0.960784,0.960784,0.960784}
  \definecolor{gainsboro}{rgb}{0.862745,0.862745,0.862745}
  \definecolor{floral white}{rgb}{1.000000,0.980392,0.941176}
  \definecolor{FloralWhite}{rgb}{1.000000,0.980392,0.941176}
  \definecolor{old lace}{rgb}{0.992157,0.960784,0.901961}
  \definecolor{OldLace}{rgb}{0.992157,0.960784,0.901961}
  \definecolor{linen}{rgb}{0.980392,0.941176,0.901961}
  \definecolor{antique white}{rgb}{0.980392,0.921569,0.843137}
  \definecolor{AntiqueWhite}{rgb}{0.980392,0.921569,0.843137}
  \definecolor{papaya whip}{rgb}{1.000000,0.937255,0.835294}
  \definecolor{PapayaWhip}{rgb}{1.000000,0.937255,0.835294}
  \definecolor{blanched almond}{rgb}{1.000000,0.921569,0.803922}
  \definecolor{BlanchedAlmond}{rgb}{1.000000,0.921569,0.803922}
  \definecolor{bisque}{rgb}{1.000000,0.894118,0.768627}
  \definecolor{peach puff}{rgb}{1.000000,0.854902,0.725490}
  \definecolor{PeachPuff}{rgb}{1.000000,0.854902,0.725490}
  \definecolor{navajo white}{rgb}{1.000000,0.870588,0.678431}
  \definecolor{NavajoWhite}{rgb}{1.000000,0.870588,0.678431}
  \definecolor{moccasin}{rgb}{1.000000,0.894118,0.709804}
  \definecolor{cornsilk}{rgb}{1.000000,0.972549,0.862745}
  \definecolor{ivory}{rgb}{1.000000,1.000000,0.941176}
  \definecolor{lemon chiffon}{rgb}{1.000000,0.980392,0.803922}
  \definecolor{LemonChiffon}{rgb}{1.000000,0.980392,0.803922}
  \definecolor{seashell}{rgb}{1.000000,0.960784,0.933333}
  \definecolor{honeydew}{rgb}{0.941176,1.000000,0.941176}
  \definecolor{mint cream}{rgb}{0.960784,1.000000,0.980392}
  \definecolor{MintCream}{rgb}{0.960784,1.000000,0.980392}
  \definecolor{azure}{rgb}{0.941176,1.000000,1.000000}
  \definecolor{alice blue}{rgb}{0.941176,0.972549,1.000000}
  \definecolor{AliceBlue}{rgb}{0.941176,0.972549,1.000000}
  \definecolor{lavender}{rgb}{0.901961,0.901961,0.980392}
  \definecolor{lavender blush}{rgb}{1.000000,0.941176,0.960784}
  \definecolor{LavenderBlush}{rgb}{1.000000,0.941176,0.960784}
  \definecolor{misty rose}{rgb}{1.000000,0.894118,0.882353}
  \definecolor{MistyRose}{rgb}{1.000000,0.894118,0.882353}
  \definecolor{white}{rgb}{1.000000,1.000000,1.000000}
  \definecolor{black}{rgb}{0.000000,0.000000,0.000000}
  \definecolor{dark slate gray}{rgb}{0.184314,0.309804,0.309804}
  \definecolor{DarkSlateGray}{rgb}{0.184314,0.309804,0.309804}
  \definecolor{dark slate grey}{rgb}{0.184314,0.309804,0.309804}
  \definecolor{DarkSlateGrey}{rgb}{0.184314,0.309804,0.309804}
  \definecolor{dim gray}{rgb}{0.411765,0.411765,0.411765}
  \definecolor{DimGray}{rgb}{0.411765,0.411765,0.411765}
  \definecolor{dim grey}{rgb}{0.411765,0.411765,0.411765}
  \definecolor{DimGrey}{rgb}{0.411765,0.411765,0.411765}
  \definecolor{slate gray}{rgb}{0.439216,0.501961,0.564706}
  \definecolor{SlateGray}{rgb}{0.439216,0.501961,0.564706}
  \definecolor{slate grey}{rgb}{0.439216,0.501961,0.564706}
  \definecolor{SlateGrey}{rgb}{0.439216,0.501961,0.564706}
  \definecolor{light slate gray}{rgb}{0.466667,0.533333,0.600000}
  \definecolor{LightSlateGray}{rgb}{0.466667,0.533333,0.600000}
  \definecolor{light slate grey}{rgb}{0.466667,0.533333,0.600000}
  \definecolor{LightSlateGrey}{rgb}{0.466667,0.533333,0.600000}
  \definecolor{gray}{rgb}{0.745098,0.745098,0.745098}
  \definecolor{grey}{rgb}{0.745098,0.745098,0.745098}
  \definecolor{light grey}{rgb}{0.827451,0.827451,0.827451}
  \definecolor{LightGrey}{rgb}{0.827451,0.827451,0.827451}
  \definecolor{light gray}{rgb}{0.827451,0.827451,0.827451}
  \definecolor{LightGray}{rgb}{0.827451,0.827451,0.827451}
  \definecolor{midnight blue}{rgb}{0.098039,0.098039,0.439216}
  \definecolor{MidnightBlue}{rgb}{0.098039,0.098039,0.439216}
  \definecolor{navy}{rgb}{0.000000,0.000000,0.501961}
  \definecolor{navy blue}{rgb}{0.000000,0.000000,0.501961}
  \definecolor{NavyBlue}{rgb}{0.000000,0.000000,0.501961}
  \definecolor{cornflower blue}{rgb}{0.392157,0.584314,0.929412}
  \definecolor{CornflowerBlue}{rgb}{0.392157,0.584314,0.929412}
  \definecolor{dark slate blue}{rgb}{0.282353,0.239216,0.545098}
  \definecolor{DarkSlateBlue}{rgb}{0.282353,0.239216,0.545098}
  \definecolor{slate blue}{rgb}{0.415686,0.352941,0.803922}
  \definecolor{SlateBlue}{rgb}{0.415686,0.352941,0.803922}
  \definecolor{medium slate blue}{rgb}{0.482353,0.407843,0.933333}
  \definecolor{MediumSlateBlue}{rgb}{0.482353,0.407843,0.933333}
  \definecolor{light slate blue}{rgb}{0.517647,0.439216,1.000000}
  \definecolor{LightSlateBlue}{rgb}{0.517647,0.439216,1.000000}
  \definecolor{medium blue}{rgb}{0.000000,0.000000,0.803922}
  \definecolor{MediumBlue}{rgb}{0.000000,0.000000,0.803922}
  \definecolor{royal blue}{rgb}{0.254902,0.411765,0.882353}
  \definecolor{RoyalBlue}{rgb}{0.254902,0.411765,0.882353}
  \definecolor{blue}{rgb}{0.000000,0.000000,1.000000}
  \definecolor{dodger blue}{rgb}{0.117647,0.564706,1.000000}
  \definecolor{DodgerBlue}{rgb}{0.117647,0.564706,1.000000}
  \definecolor{deep sky blue}{rgb}{0.000000,0.749020,1.000000}
  \definecolor{DeepSkyBlue}{rgb}{0.000000,0.749020,1.000000}
  \definecolor{sky blue}{rgb}{0.529412,0.807843,0.921569}
  \definecolor{SkyBlue}{rgb}{0.529412,0.807843,0.921569}
  \definecolor{light sky blue}{rgb}{0.529412,0.807843,0.980392}
  \definecolor{LightSkyBlue}{rgb}{0.529412,0.807843,0.980392}
  \definecolor{steel blue}{rgb}{0.274510,0.509804,0.705882}
  \definecolor{SteelBlue}{rgb}{0.274510,0.509804,0.705882}
  \definecolor{light steel blue}{rgb}{0.690196,0.768627,0.870588}
  \definecolor{LightSteelBlue}{rgb}{0.690196,0.768627,0.870588}
  \definecolor{light blue}{rgb}{0.678431,0.847059,0.901961}
  \definecolor{LightBlue}{rgb}{0.678431,0.847059,0.901961}
  \definecolor{powder blue}{rgb}{0.690196,0.878431,0.901961}
  \definecolor{PowderBlue}{rgb}{0.690196,0.878431,0.901961}
  \definecolor{pale turquoise}{rgb}{0.686275,0.933333,0.933333}
  \definecolor{PaleTurquoise}{rgb}{0.686275,0.933333,0.933333}
  \definecolor{dark turquoise}{rgb}{0.000000,0.807843,0.819608}
  \definecolor{DarkTurquoise}{rgb}{0.000000,0.807843,0.819608}
  \definecolor{medium turquoise}{rgb}{0.282353,0.819608,0.800000}
  \definecolor{MediumTurquoise}{rgb}{0.282353,0.819608,0.800000}
  \definecolor{turquoise}{rgb}{0.250980,0.878431,0.815686}
  \definecolor{cyan}{rgb}{0.000000,1.000000,1.000000}
  \definecolor{light cyan}{rgb}{0.878431,1.000000,1.000000}
  \definecolor{LightCyan}{rgb}{0.878431,1.000000,1.000000}
  \definecolor{cadet blue}{rgb}{0.372549,0.619608,0.627451}
  \definecolor{CadetBlue}{rgb}{0.372549,0.619608,0.627451}
  \definecolor{medium aquamarine}{rgb}{0.400000,0.803922,0.666667}
  \definecolor{MediumAquamarine}{rgb}{0.400000,0.803922,0.666667}
  \definecolor{aquamarine}{rgb}{0.498039,1.000000,0.831373}
  \definecolor{dark green}{rgb}{0.000000,0.392157,0.000000}
  \definecolor{DarkGreen}{rgb}{0.000000,0.392157,0.000000}
  \definecolor{dark olive green}{rgb}{0.333333,0.419608,0.184314}
  \definecolor{DarkOliveGreen}{rgb}{0.333333,0.419608,0.184314}
  \definecolor{dark sea green}{rgb}{0.560784,0.737255,0.560784}
  \definecolor{DarkSeaGreen}{rgb}{0.560784,0.737255,0.560784}
  \definecolor{sea green}{rgb}{0.180392,0.545098,0.341176}
  \definecolor{SeaGreen}{rgb}{0.180392,0.545098,0.341176}
  \definecolor{medium sea green}{rgb}{0.235294,0.701961,0.443137}
  \definecolor{MediumSeaGreen}{rgb}{0.235294,0.701961,0.443137}
  \definecolor{light sea green}{rgb}{0.125490,0.698039,0.666667}
  \definecolor{LightSeaGreen}{rgb}{0.125490,0.698039,0.666667}
  \definecolor{pale green}{rgb}{0.596078,0.984314,0.596078}
  \definecolor{PaleGreen}{rgb}{0.596078,0.984314,0.596078}
  \definecolor{spring green}{rgb}{0.000000,1.000000,0.498039}
  \definecolor{SpringGreen}{rgb}{0.000000,1.000000,0.498039}
  \definecolor{lawn green}{rgb}{0.486275,0.988235,0.000000}
  \definecolor{LawnGreen}{rgb}{0.486275,0.988235,0.000000}
  \definecolor{green}{rgb}{0.000000,1.000000,0.000000}
  \definecolor{chartreuse}{rgb}{0.498039,1.000000,0.000000}
  \definecolor{medium spring green}{rgb}{0.000000,0.980392,0.603922}
  \definecolor{MediumSpringGreen}{rgb}{0.000000,0.980392,0.603922}
  \definecolor{green yellow}{rgb}{0.678431,1.000000,0.184314}
  \definecolor{GreenYellow}{rgb}{0.678431,1.000000,0.184314}
  \definecolor{lime green}{rgb}{0.196078,0.803922,0.196078}
  \definecolor{LimeGreen}{rgb}{0.196078,0.803922,0.196078}
  \definecolor{yellow green}{rgb}{0.603922,0.803922,0.196078}
  \definecolor{YellowGreen}{rgb}{0.603922,0.803922,0.196078}
  \definecolor{forest green}{rgb}{0.133333,0.545098,0.133333}
  \definecolor{ForestGreen}{rgb}{0.133333,0.545098,0.133333}
  \definecolor{olive drab}{rgb}{0.419608,0.556863,0.137255}
  \definecolor{OliveDrab}{rgb}{0.419608,0.556863,0.137255}
  \definecolor{dark khaki}{rgb}{0.741176,0.717647,0.419608}
  \definecolor{DarkKhaki}{rgb}{0.741176,0.717647,0.419608}
  \definecolor{khaki}{rgb}{0.941176,0.901961,0.549020}
  \definecolor{pale goldenrod}{rgb}{0.933333,0.909804,0.666667}
  \definecolor{PaleGoldenrod}{rgb}{0.933333,0.909804,0.666667}
  \definecolor{light goldenrod yellow}{rgb}{0.980392,0.980392,0.823529}
  \definecolor{LightGoldenrodYellow}{rgb}{0.980392,0.980392,0.823529}
  \definecolor{light yellow}{rgb}{1.000000,1.000000,0.878431}
  \definecolor{LightYellow}{rgb}{1.000000,1.000000,0.878431}
  \definecolor{yellow}{rgb}{1.000000,1.000000,0.000000}
  \definecolor{gold}{rgb}{1.000000,0.843137,0.000000}
  \definecolor{light goldenrod}{rgb}{0.933333,0.866667,0.509804}
  \definecolor{LightGoldenrod}{rgb}{0.933333,0.866667,0.509804}
  \definecolor{goldenrod}{rgb}{0.854902,0.647059,0.125490}
  \definecolor{dark goldenrod}{rgb}{0.721569,0.525490,0.043137}
  \definecolor{DarkGoldenrod}{rgb}{0.721569,0.525490,0.043137}
  \definecolor{rosy brown}{rgb}{0.737255,0.560784,0.560784}
  \definecolor{RosyBrown}{rgb}{0.737255,0.560784,0.560784}
  \definecolor{indian red}{rgb}{0.803922,0.360784,0.360784}
  \definecolor{IndianRed}{rgb}{0.803922,0.360784,0.360784}
  \definecolor{saddle brown}{rgb}{0.545098,0.270588,0.074510}
  \definecolor{SaddleBrown}{rgb}{0.545098,0.270588,0.074510}
  \definecolor{sienna}{rgb}{0.627451,0.321569,0.176471}
  \definecolor{peru}{rgb}{0.803922,0.521569,0.247059}
  \definecolor{burlywood}{rgb}{0.870588,0.721569,0.529412}
  \definecolor{beige}{rgb}{0.960784,0.960784,0.862745}
  \definecolor{wheat}{rgb}{0.960784,0.870588,0.701961}
  \definecolor{sandy brown}{rgb}{0.956863,0.643137,0.376471}
  \definecolor{SandyBrown}{rgb}{0.956863,0.643137,0.376471}
  \definecolor{tan}{rgb}{0.823529,0.705882,0.549020}
  \definecolor{chocolate}{rgb}{0.823529,0.411765,0.117647}
  \definecolor{firebrick}{rgb}{0.698039,0.133333,0.133333}
  \definecolor{brown}{rgb}{0.647059,0.164706,0.164706}
  \definecolor{dark salmon}{rgb}{0.913725,0.588235,0.478431}
  \definecolor{DarkSalmon}{rgb}{0.913725,0.588235,0.478431}
  \definecolor{salmon}{rgb}{0.980392,0.501961,0.447059}
  \definecolor{light salmon}{rgb}{1.000000,0.627451,0.478431}
  \definecolor{LightSalmon}{rgb}{1.000000,0.627451,0.478431}
  \definecolor{orange}{rgb}{1.000000,0.647059,0.000000}
  \definecolor{dark orange}{rgb}{1.000000,0.549020,0.000000}
  \definecolor{DarkOrange}{rgb}{1.000000,0.549020,0.000000}
  \definecolor{coral}{rgb}{1.000000,0.498039,0.313726}
  \definecolor{light coral}{rgb}{0.941176,0.501961,0.501961}
  \definecolor{LightCoral}{rgb}{0.941176,0.501961,0.501961}
  \definecolor{tomato}{rgb}{1.000000,0.388235,0.278431}
  \definecolor{orange red}{rgb}{1.000000,0.270588,0.000000}
  \definecolor{OrangeRed}{rgb}{1.000000,0.270588,0.000000}
  \definecolor{red}{rgb}{1.000000,0.000000,0.000000}
  \definecolor{hot pink}{rgb}{1.000000,0.411765,0.705882}
  \definecolor{HotPink}{rgb}{1.000000,0.411765,0.705882}
  \definecolor{deep pink}{rgb}{1.000000,0.078431,0.576471}
  \definecolor{DeepPink}{rgb}{1.000000,0.078431,0.576471}
  \definecolor{pink}{rgb}{1.000000,0.752941,0.796078}
  \definecolor{light pink}{rgb}{1.000000,0.713726,0.756863}
  \definecolor{LightPink}{rgb}{1.000000,0.713726,0.756863}
  \definecolor{pale violet red}{rgb}{0.858824,0.439216,0.576471}
  \definecolor{PaleVioletRed}{rgb}{0.858824,0.439216,0.576471}
  \definecolor{maroon}{rgb}{0.690196,0.188235,0.376471}
  \definecolor{medium violet red}{rgb}{0.780392,0.082353,0.521569}
  \definecolor{MediumVioletRed}{rgb}{0.780392,0.082353,0.521569}
  \definecolor{violet red}{rgb}{0.815686,0.125490,0.564706}
  \definecolor{VioletRed}{rgb}{0.815686,0.125490,0.564706}
  \definecolor{magenta}{rgb}{1.000000,0.000000,1.000000}
  \definecolor{violet}{rgb}{0.933333,0.509804,0.933333}
  \definecolor{plum}{rgb}{0.866667,0.627451,0.866667}
  \definecolor{orchid}{rgb}{0.854902,0.439216,0.839216}
  \definecolor{medium orchid}{rgb}{0.729412,0.333333,0.827451}
  \definecolor{MediumOrchid}{rgb}{0.729412,0.333333,0.827451}
  \definecolor{dark orchid}{rgb}{0.600000,0.196078,0.800000}
  \definecolor{DarkOrchid}{rgb}{0.600000,0.196078,0.800000}
  \definecolor{dark violet}{rgb}{0.580392,0.000000,0.827451}
  \definecolor{DarkViolet}{rgb}{0.580392,0.000000,0.827451}
  \definecolor{blue violet}{rgb}{0.541176,0.168627,0.886275}
  \definecolor{BlueViolet}{rgb}{0.541176,0.168627,0.886275}
  \definecolor{purple}{rgb}{0.627451,0.125490,0.941176}
  \definecolor{medium purple}{rgb}{0.576471,0.439216,0.858824}
  \definecolor{MediumPurple}{rgb}{0.576471,0.439216,0.858824}
  \definecolor{thistle}{rgb}{0.847059,0.749020,0.847059}
  \definecolor{snow1}{rgb}{1.000000,0.980392,0.980392}
  \definecolor{snow2}{rgb}{0.933333,0.913725,0.913725}
  \definecolor{snow3}{rgb}{0.803922,0.788235,0.788235}
  \definecolor{snow4}{rgb}{0.545098,0.537255,0.537255}
  \definecolor{seashell1}{rgb}{1.000000,0.960784,0.933333}
  \definecolor{seashell2}{rgb}{0.933333,0.898039,0.870588}
  \definecolor{seashell3}{rgb}{0.803922,0.772549,0.749020}
  \definecolor{seashell4}{rgb}{0.545098,0.525490,0.509804}
  \definecolor{AntiqueWhite1}{rgb}{1.000000,0.937255,0.858824}
  \definecolor{AntiqueWhite2}{rgb}{0.933333,0.874510,0.800000}
  \definecolor{AntiqueWhite3}{rgb}{0.803922,0.752941,0.690196}
  \definecolor{AntiqueWhite4}{rgb}{0.545098,0.513726,0.470588}
  \definecolor{bisque1}{rgb}{1.000000,0.894118,0.768627}
  \definecolor{bisque2}{rgb}{0.933333,0.835294,0.717647}
  \definecolor{bisque3}{rgb}{0.803922,0.717647,0.619608}
  \definecolor{bisque4}{rgb}{0.545098,0.490196,0.419608}
  \definecolor{PeachPuff1}{rgb}{1.000000,0.854902,0.725490}
  \definecolor{PeachPuff2}{rgb}{0.933333,0.796078,0.678431}
  \definecolor{PeachPuff3}{rgb}{0.803922,0.686275,0.584314}
  \definecolor{PeachPuff4}{rgb}{0.545098,0.466667,0.396078}
  \definecolor{NavajoWhite1}{rgb}{1.000000,0.870588,0.678431}
  \definecolor{NavajoWhite2}{rgb}{0.933333,0.811765,0.631373}
  \definecolor{NavajoWhite3}{rgb}{0.803922,0.701961,0.545098}
  \definecolor{NavajoWhite4}{rgb}{0.545098,0.474510,0.368627}
  \definecolor{LemonChiffon1}{rgb}{1.000000,0.980392,0.803922}
  \definecolor{LemonChiffon2}{rgb}{0.933333,0.913725,0.749020}
  \definecolor{LemonChiffon3}{rgb}{0.803922,0.788235,0.647059}
  \definecolor{LemonChiffon4}{rgb}{0.545098,0.537255,0.439216}
  \definecolor{cornsilk1}{rgb}{1.000000,0.972549,0.862745}
  \definecolor{cornsilk2}{rgb}{0.933333,0.909804,0.803922}
  \definecolor{cornsilk3}{rgb}{0.803922,0.784314,0.694118}
  \definecolor{cornsilk4}{rgb}{0.545098,0.533333,0.470588}
  \definecolor{ivory1}{rgb}{1.000000,1.000000,0.941176}
  \definecolor{ivory2}{rgb}{0.933333,0.933333,0.878431}
  \definecolor{ivory3}{rgb}{0.803922,0.803922,0.756863}
  \definecolor{ivory4}{rgb}{0.545098,0.545098,0.513726}
  \definecolor{honeydew1}{rgb}{0.941176,1.000000,0.941176}
  \definecolor{honeydew2}{rgb}{0.878431,0.933333,0.878431}
  \definecolor{honeydew3}{rgb}{0.756863,0.803922,0.756863}
  \definecolor{honeydew4}{rgb}{0.513726,0.545098,0.513726}
  \definecolor{LavenderBlush1}{rgb}{1.000000,0.941176,0.960784}
  \definecolor{LavenderBlush2}{rgb}{0.933333,0.878431,0.898039}
  \definecolor{LavenderBlush3}{rgb}{0.803922,0.756863,0.772549}
  \definecolor{LavenderBlush4}{rgb}{0.545098,0.513726,0.525490}
  \definecolor{MistyRose1}{rgb}{1.000000,0.894118,0.882353}
  \definecolor{MistyRose2}{rgb}{0.933333,0.835294,0.823529}
  \definecolor{MistyRose3}{rgb}{0.803922,0.717647,0.709804}
  \definecolor{MistyRose4}{rgb}{0.545098,0.490196,0.482353}
  \definecolor{azure1}{rgb}{0.941176,1.000000,1.000000}
  \definecolor{azure2}{rgb}{0.878431,0.933333,0.933333}
  \definecolor{azure3}{rgb}{0.756863,0.803922,0.803922}
  \definecolor{azure4}{rgb}{0.513726,0.545098,0.545098}
  \definecolor{SlateBlue1}{rgb}{0.513726,0.435294,1.000000}
  \definecolor{SlateBlue2}{rgb}{0.478431,0.403922,0.933333}
  \definecolor{SlateBlue3}{rgb}{0.411765,0.349020,0.803922}
  \definecolor{SlateBlue4}{rgb}{0.278431,0.235294,0.545098}
  \definecolor{RoyalBlue1}{rgb}{0.282353,0.462745,1.000000}
  \definecolor{RoyalBlue2}{rgb}{0.262745,0.431373,0.933333}
  \definecolor{RoyalBlue3}{rgb}{0.227451,0.372549,0.803922}
  \definecolor{RoyalBlue4}{rgb}{0.152941,0.250980,0.545098}
  \definecolor{blue1}{rgb}{0.000000,0.000000,1.000000}
  \definecolor{blue2}{rgb}{0.000000,0.000000,0.933333}
  \definecolor{blue3}{rgb}{0.000000,0.000000,0.803922}
  \definecolor{blue4}{rgb}{0.000000,0.000000,0.545098}
  \definecolor{DodgerBlue1}{rgb}{0.117647,0.564706,1.000000}
  \definecolor{DodgerBlue2}{rgb}{0.109804,0.525490,0.933333}
  \definecolor{DodgerBlue3}{rgb}{0.094118,0.454902,0.803922}
  \definecolor{DodgerBlue4}{rgb}{0.062745,0.305882,0.545098}
  \definecolor{SteelBlue1}{rgb}{0.388235,0.721569,1.000000}
  \definecolor{SteelBlue2}{rgb}{0.360784,0.674510,0.933333}
  \definecolor{SteelBlue3}{rgb}{0.309804,0.580392,0.803922}
  \definecolor{SteelBlue4}{rgb}{0.211765,0.392157,0.545098}
  \definecolor{DeepSkyBlue1}{rgb}{0.000000,0.749020,1.000000}
  \definecolor{DeepSkyBlue2}{rgb}{0.000000,0.698039,0.933333}
  \definecolor{DeepSkyBlue3}{rgb}{0.000000,0.603922,0.803922}
  \definecolor{DeepSkyBlue4}{rgb}{0.000000,0.407843,0.545098}
  \definecolor{SkyBlue1}{rgb}{0.529412,0.807843,1.000000}
  \definecolor{SkyBlue2}{rgb}{0.494118,0.752941,0.933333}
  \definecolor{SkyBlue3}{rgb}{0.423529,0.650980,0.803922}
  \definecolor{SkyBlue4}{rgb}{0.290196,0.439216,0.545098}
  \definecolor{LightSkyBlue1}{rgb}{0.690196,0.886275,1.000000}
  \definecolor{LightSkyBlue2}{rgb}{0.643137,0.827451,0.933333}
  \definecolor{LightSkyBlue3}{rgb}{0.552941,0.713726,0.803922}
  \definecolor{LightSkyBlue4}{rgb}{0.376471,0.482353,0.545098}
  \definecolor{SlateGray1}{rgb}{0.776471,0.886275,1.000000}
  \definecolor{SlateGray2}{rgb}{0.725490,0.827451,0.933333}
  \definecolor{SlateGray3}{rgb}{0.623529,0.713726,0.803922}
  \definecolor{SlateGray4}{rgb}{0.423529,0.482353,0.545098}
  \definecolor{LightSteelBlue1}{rgb}{0.792157,0.882353,1.000000}
  \definecolor{LightSteelBlue2}{rgb}{0.737255,0.823529,0.933333}
  \definecolor{LightSteelBlue3}{rgb}{0.635294,0.709804,0.803922}
  \definecolor{LightSteelBlue4}{rgb}{0.431373,0.482353,0.545098}
  \definecolor{LightBlue1}{rgb}{0.749020,0.937255,1.000000}
  \definecolor{LightBlue2}{rgb}{0.698039,0.874510,0.933333}
  \definecolor{LightBlue3}{rgb}{0.603922,0.752941,0.803922}
  \definecolor{LightBlue4}{rgb}{0.407843,0.513726,0.545098}
  \definecolor{LightCyan1}{rgb}{0.878431,1.000000,1.000000}
  \definecolor{LightCyan2}{rgb}{0.819608,0.933333,0.933333}
  \definecolor{LightCyan3}{rgb}{0.705882,0.803922,0.803922}
  \definecolor{LightCyan4}{rgb}{0.478431,0.545098,0.545098}
  \definecolor{PaleTurquoise1}{rgb}{0.733333,1.000000,1.000000}
  \definecolor{PaleTurquoise2}{rgb}{0.682353,0.933333,0.933333}
  \definecolor{PaleTurquoise3}{rgb}{0.588235,0.803922,0.803922}
  \definecolor{PaleTurquoise4}{rgb}{0.400000,0.545098,0.545098}
  \definecolor{CadetBlue1}{rgb}{0.596078,0.960784,1.000000}
  \definecolor{CadetBlue2}{rgb}{0.556863,0.898039,0.933333}
  \definecolor{CadetBlue3}{rgb}{0.478431,0.772549,0.803922}
  \definecolor{CadetBlue4}{rgb}{0.325490,0.525490,0.545098}
  \definecolor{turquoise1}{rgb}{0.000000,0.960784,1.000000}
  \definecolor{turquoise2}{rgb}{0.000000,0.898039,0.933333}
  \definecolor{turquoise3}{rgb}{0.000000,0.772549,0.803922}
  \definecolor{turquoise4}{rgb}{0.000000,0.525490,0.545098}
  \definecolor{cyan1}{rgb}{0.000000,1.000000,1.000000}
  \definecolor{cyan2}{rgb}{0.000000,0.933333,0.933333}
  \definecolor{cyan3}{rgb}{0.000000,0.803922,0.803922}
  \definecolor{cyan4}{rgb}{0.000000,0.545098,0.545098}
  \definecolor{DarkSlateGray1}{rgb}{0.592157,1.000000,1.000000}
  \definecolor{DarkSlateGray2}{rgb}{0.552941,0.933333,0.933333}
  \definecolor{DarkSlateGray3}{rgb}{0.474510,0.803922,0.803922}
  \definecolor{DarkSlateGray4}{rgb}{0.321569,0.545098,0.545098}
  \definecolor{aquamarine1}{rgb}{0.498039,1.000000,0.831373}
  \definecolor{aquamarine2}{rgb}{0.462745,0.933333,0.776471}
  \definecolor{aquamarine3}{rgb}{0.400000,0.803922,0.666667}
  \definecolor{aquamarine4}{rgb}{0.270588,0.545098,0.454902}
  \definecolor{DarkSeaGreen1}{rgb}{0.756863,1.000000,0.756863}
  \definecolor{DarkSeaGreen2}{rgb}{0.705882,0.933333,0.705882}
  \definecolor{DarkSeaGreen3}{rgb}{0.607843,0.803922,0.607843}
  \definecolor{DarkSeaGreen4}{rgb}{0.411765,0.545098,0.411765}
  \definecolor{SeaGreen1}{rgb}{0.329412,1.000000,0.623529}
  \definecolor{SeaGreen2}{rgb}{0.305882,0.933333,0.580392}
  \definecolor{SeaGreen3}{rgb}{0.262745,0.803922,0.501961}
  \definecolor{SeaGreen4}{rgb}{0.180392,0.545098,0.341176}
  \definecolor{PaleGreen1}{rgb}{0.603922,1.000000,0.603922}
  \definecolor{PaleGreen2}{rgb}{0.564706,0.933333,0.564706}
  \definecolor{PaleGreen3}{rgb}{0.486275,0.803922,0.486275}
  \definecolor{PaleGreen4}{rgb}{0.329412,0.545098,0.329412}
  \definecolor{SpringGreen1}{rgb}{0.000000,1.000000,0.498039}
  \definecolor{SpringGreen2}{rgb}{0.000000,0.933333,0.462745}
  \definecolor{SpringGreen3}{rgb}{0.000000,0.803922,0.400000}
  \definecolor{SpringGreen4}{rgb}{0.000000,0.545098,0.270588}
  \definecolor{green1}{rgb}{0.000000,1.000000,0.000000}
  \definecolor{green2}{rgb}{0.000000,0.933333,0.000000}
  \definecolor{green3}{rgb}{0.000000,0.803922,0.000000}
  \definecolor{green4}{rgb}{0.000000,0.545098,0.000000}
  \definecolor{chartreuse1}{rgb}{0.498039,1.000000,0.000000}
  \definecolor{chartreuse2}{rgb}{0.462745,0.933333,0.000000}
  \definecolor{chartreuse3}{rgb}{0.400000,0.803922,0.000000}
  \definecolor{chartreuse4}{rgb}{0.270588,0.545098,0.000000}
  \definecolor{OliveDrab1}{rgb}{0.752941,1.000000,0.243137}
  \definecolor{OliveDrab2}{rgb}{0.701961,0.933333,0.227451}
  \definecolor{OliveDrab3}{rgb}{0.603922,0.803922,0.196078}
  \definecolor{OliveDrab4}{rgb}{0.411765,0.545098,0.133333}
  \definecolor{DarkOliveGreen1}{rgb}{0.792157,1.000000,0.439216}
  \definecolor{DarkOliveGreen2}{rgb}{0.737255,0.933333,0.407843}
  \definecolor{DarkOliveGreen3}{rgb}{0.635294,0.803922,0.352941}
  \definecolor{DarkOliveGreen4}{rgb}{0.431373,0.545098,0.239216}
  \definecolor{khaki1}{rgb}{1.000000,0.964706,0.560784}
  \definecolor{khaki2}{rgb}{0.933333,0.901961,0.521569}
  \definecolor{khaki3}{rgb}{0.803922,0.776471,0.450980}
  \definecolor{khaki4}{rgb}{0.545098,0.525490,0.305882}
  \definecolor{LightGoldenrod1}{rgb}{1.000000,0.925490,0.545098}
  \definecolor{LightGoldenrod2}{rgb}{0.933333,0.862745,0.509804}
  \definecolor{LightGoldenrod3}{rgb}{0.803922,0.745098,0.439216}
  \definecolor{LightGoldenrod4}{rgb}{0.545098,0.505882,0.298039}
  \definecolor{LightYellow1}{rgb}{1.000000,1.000000,0.878431}
  \definecolor{LightYellow2}{rgb}{0.933333,0.933333,0.819608}
  \definecolor{LightYellow3}{rgb}{0.803922,0.803922,0.705882}
  \definecolor{LightYellow4}{rgb}{0.545098,0.545098,0.478431}
  \definecolor{yellow1}{rgb}{1.000000,1.000000,0.000000}
  \definecolor{yellow2}{rgb}{0.933333,0.933333,0.000000}
  \definecolor{yellow3}{rgb}{0.803922,0.803922,0.000000}
  \definecolor{yellow4}{rgb}{0.545098,0.545098,0.000000}
  \definecolor{gold1}{rgb}{1.000000,0.843137,0.000000}
  \definecolor{gold2}{rgb}{0.933333,0.788235,0.000000}
  \definecolor{gold3}{rgb}{0.803922,0.678431,0.000000}
  \definecolor{gold4}{rgb}{0.545098,0.458824,0.000000}
  \definecolor{goldenrod1}{rgb}{1.000000,0.756863,0.145098}
  \definecolor{goldenrod2}{rgb}{0.933333,0.705882,0.133333}
  \definecolor{goldenrod3}{rgb}{0.803922,0.607843,0.113725}
  \definecolor{goldenrod4}{rgb}{0.545098,0.411765,0.078431}
  \definecolor{DarkGoldenrod1}{rgb}{1.000000,0.725490,0.058824}
  \definecolor{DarkGoldenrod2}{rgb}{0.933333,0.678431,0.054902}
  \definecolor{DarkGoldenrod3}{rgb}{0.803922,0.584314,0.047059}
  \definecolor{DarkGoldenrod4}{rgb}{0.545098,0.396078,0.031373}
  \definecolor{RosyBrown1}{rgb}{1.000000,0.756863,0.756863}
  \definecolor{RosyBrown2}{rgb}{0.933333,0.705882,0.705882}
  \definecolor{RosyBrown3}{rgb}{0.803922,0.607843,0.607843}
  \definecolor{RosyBrown4}{rgb}{0.545098,0.411765,0.411765}
  \definecolor{IndianRed1}{rgb}{1.000000,0.415686,0.415686}
  \definecolor{IndianRed2}{rgb}{0.933333,0.388235,0.388235}
  \definecolor{IndianRed3}{rgb}{0.803922,0.333333,0.333333}
  \definecolor{IndianRed4}{rgb}{0.545098,0.227451,0.227451}
  \definecolor{sienna1}{rgb}{1.000000,0.509804,0.278431}
  \definecolor{sienna2}{rgb}{0.933333,0.474510,0.258824}
  \definecolor{sienna3}{rgb}{0.803922,0.407843,0.223529}
  \definecolor{sienna4}{rgb}{0.545098,0.278431,0.149020}
  \definecolor{burlywood1}{rgb}{1.000000,0.827451,0.607843}
  \definecolor{burlywood2}{rgb}{0.933333,0.772549,0.568627}
  \definecolor{burlywood3}{rgb}{0.803922,0.666667,0.490196}
  \definecolor{burlywood4}{rgb}{0.545098,0.450980,0.333333}
  \definecolor{wheat1}{rgb}{1.000000,0.905882,0.729412}
  \definecolor{wheat2}{rgb}{0.933333,0.847059,0.682353}
  \definecolor{wheat3}{rgb}{0.803922,0.729412,0.588235}
  \definecolor{wheat4}{rgb}{0.545098,0.494118,0.400000}
  \definecolor{tan1}{rgb}{1.000000,0.647059,0.309804}
  \definecolor{tan2}{rgb}{0.933333,0.603922,0.286275}
  \definecolor{tan3}{rgb}{0.803922,0.521569,0.247059}
  \definecolor{tan4}{rgb}{0.545098,0.352941,0.168627}
  \definecolor{chocolate1}{rgb}{1.000000,0.498039,0.141176}
  \definecolor{chocolate2}{rgb}{0.933333,0.462745,0.129412}
  \definecolor{chocolate3}{rgb}{0.803922,0.400000,0.113725}
  \definecolor{chocolate4}{rgb}{0.545098,0.270588,0.074510}
  \definecolor{firebrick1}{rgb}{1.000000,0.188235,0.188235}
  \definecolor{firebrick2}{rgb}{0.933333,0.172549,0.172549}
  \definecolor{firebrick3}{rgb}{0.803922,0.149020,0.149020}
  \definecolor{firebrick4}{rgb}{0.545098,0.101961,0.101961}
  \definecolor{brown1}{rgb}{1.000000,0.250980,0.250980}
  \definecolor{brown2}{rgb}{0.933333,0.231373,0.231373}
  \definecolor{brown3}{rgb}{0.803922,0.200000,0.200000}
  \definecolor{brown4}{rgb}{0.545098,0.137255,0.137255}
  \definecolor{salmon1}{rgb}{1.000000,0.549020,0.411765}
  \definecolor{salmon2}{rgb}{0.933333,0.509804,0.384314}
  \definecolor{salmon3}{rgb}{0.803922,0.439216,0.329412}
  \definecolor{salmon4}{rgb}{0.545098,0.298039,0.223529}
  \definecolor{LightSalmon1}{rgb}{1.000000,0.627451,0.478431}
  \definecolor{LightSalmon2}{rgb}{0.933333,0.584314,0.447059}
  \definecolor{LightSalmon3}{rgb}{0.803922,0.505882,0.384314}
  \definecolor{LightSalmon4}{rgb}{0.545098,0.341176,0.258824}
  \definecolor{orange1}{rgb}{1.000000,0.647059,0.000000}
  \definecolor{orange2}{rgb}{0.933333,0.603922,0.000000}
  \definecolor{orange3}{rgb}{0.803922,0.521569,0.000000}
  \definecolor{orange4}{rgb}{0.545098,0.352941,0.000000}
  \definecolor{DarkOrange1}{rgb}{1.000000,0.498039,0.000000}
  \definecolor{DarkOrange2}{rgb}{0.933333,0.462745,0.000000}
  \definecolor{DarkOrange3}{rgb}{0.803922,0.400000,0.000000}
  \definecolor{DarkOrange4}{rgb}{0.545098,0.270588,0.000000}
  \definecolor{coral1}{rgb}{1.000000,0.447059,0.337255}
  \definecolor{coral2}{rgb}{0.933333,0.415686,0.313726}
  \definecolor{coral3}{rgb}{0.803922,0.356863,0.270588}
  \definecolor{coral4}{rgb}{0.545098,0.243137,0.184314}
  \definecolor{tomato1}{rgb}{1.000000,0.388235,0.278431}
  \definecolor{tomato2}{rgb}{0.933333,0.360784,0.258824}
  \definecolor{tomato3}{rgb}{0.803922,0.309804,0.223529}
  \definecolor{tomato4}{rgb}{0.545098,0.211765,0.149020}
  \definecolor{OrangeRed1}{rgb}{1.000000,0.270588,0.000000}
  \definecolor{OrangeRed2}{rgb}{0.933333,0.250980,0.000000}
  \definecolor{OrangeRed3}{rgb}{0.803922,0.215686,0.000000}
  \definecolor{OrangeRed4}{rgb}{0.545098,0.145098,0.000000}
  \definecolor{red1}{rgb}{1.000000,0.000000,0.000000}
  \definecolor{red2}{rgb}{0.933333,0.000000,0.000000}
  \definecolor{red3}{rgb}{0.803922,0.000000,0.000000}
  \definecolor{red4}{rgb}{0.545098,0.000000,0.000000}
  \definecolor{DeepPink1}{rgb}{1.000000,0.078431,0.576471}
  \definecolor{DeepPink2}{rgb}{0.933333,0.070588,0.537255}
  \definecolor{DeepPink3}{rgb}{0.803922,0.062745,0.462745}
  \definecolor{DeepPink4}{rgb}{0.545098,0.039216,0.313726}
  \definecolor{HotPink1}{rgb}{1.000000,0.431373,0.705882}
  \definecolor{HotPink2}{rgb}{0.933333,0.415686,0.654902}
  \definecolor{HotPink3}{rgb}{0.803922,0.376471,0.564706}
  \definecolor{HotPink4}{rgb}{0.545098,0.227451,0.384314}
  \definecolor{pink1}{rgb}{1.000000,0.709804,0.772549}
  \definecolor{pink2}{rgb}{0.933333,0.662745,0.721569}
  \definecolor{pink3}{rgb}{0.803922,0.568627,0.619608}
  \definecolor{pink4}{rgb}{0.545098,0.388235,0.423529}
  \definecolor{LightPink1}{rgb}{1.000000,0.682353,0.725490}
  \definecolor{LightPink2}{rgb}{0.933333,0.635294,0.678431}
  \definecolor{LightPink3}{rgb}{0.803922,0.549020,0.584314}
  \definecolor{LightPink4}{rgb}{0.545098,0.372549,0.396078}
  \definecolor{PaleVioletRed1}{rgb}{1.000000,0.509804,0.670588}
  \definecolor{PaleVioletRed2}{rgb}{0.933333,0.474510,0.623529}
  \definecolor{PaleVioletRed3}{rgb}{0.803922,0.407843,0.537255}
  \definecolor{PaleVioletRed4}{rgb}{0.545098,0.278431,0.364706}
  \definecolor{maroon1}{rgb}{1.000000,0.203922,0.701961}
  \definecolor{maroon2}{rgb}{0.933333,0.188235,0.654902}
  \definecolor{maroon3}{rgb}{0.803922,0.160784,0.564706}
  \definecolor{maroon4}{rgb}{0.545098,0.109804,0.384314}
  \definecolor{VioletRed1}{rgb}{1.000000,0.243137,0.588235}
  \definecolor{VioletRed2}{rgb}{0.933333,0.227451,0.549020}
  \definecolor{VioletRed3}{rgb}{0.803922,0.196078,0.470588}
  \definecolor{VioletRed4}{rgb}{0.545098,0.133333,0.321569}
  \definecolor{magenta1}{rgb}{1.000000,0.000000,1.000000}
  \definecolor{magenta2}{rgb}{0.933333,0.000000,0.933333}
  \definecolor{magenta3}{rgb}{0.803922,0.000000,0.803922}
  \definecolor{magenta4}{rgb}{0.545098,0.000000,0.545098}
  \definecolor{orchid1}{rgb}{1.000000,0.513726,0.980392}
  \definecolor{orchid2}{rgb}{0.933333,0.478431,0.913725}
  \definecolor{orchid3}{rgb}{0.803922,0.411765,0.788235}
  \definecolor{orchid4}{rgb}{0.545098,0.278431,0.537255}
  \definecolor{plum1}{rgb}{1.000000,0.733333,1.000000}
  \definecolor{plum2}{rgb}{0.933333,0.682353,0.933333}
  \definecolor{plum3}{rgb}{0.803922,0.588235,0.803922}
  \definecolor{plum4}{rgb}{0.545098,0.400000,0.545098}
  \definecolor{MediumOrchid1}{rgb}{0.878431,0.400000,1.000000}
  \definecolor{MediumOrchid2}{rgb}{0.819608,0.372549,0.933333}
  \definecolor{MediumOrchid3}{rgb}{0.705882,0.321569,0.803922}
  \definecolor{MediumOrchid4}{rgb}{0.478431,0.215686,0.545098}
  \definecolor{DarkOrchid1}{rgb}{0.749020,0.243137,1.000000}
  \definecolor{DarkOrchid2}{rgb}{0.698039,0.227451,0.933333}
  \definecolor{DarkOrchid3}{rgb}{0.603922,0.196078,0.803922}
  \definecolor{DarkOrchid4}{rgb}{0.407843,0.133333,0.545098}
  \definecolor{purple1}{rgb}{0.607843,0.188235,1.000000}
  \definecolor{purple2}{rgb}{0.568627,0.172549,0.933333}
  \definecolor{purple3}{rgb}{0.490196,0.149020,0.803922}
  \definecolor{purple4}{rgb}{0.333333,0.101961,0.545098}
  \definecolor{MediumPurple1}{rgb}{0.670588,0.509804,1.000000}
  \definecolor{MediumPurple2}{rgb}{0.623529,0.474510,0.933333}
  \definecolor{MediumPurple3}{rgb}{0.537255,0.407843,0.803922}
  \definecolor{MediumPurple4}{rgb}{0.364706,0.278431,0.545098}
  \definecolor{thistle1}{rgb}{1.000000,0.882353,1.000000}
  \definecolor{thistle2}{rgb}{0.933333,0.823529,0.933333}
  \definecolor{thistle3}{rgb}{0.803922,0.709804,0.803922}
  \definecolor{thistle4}{rgb}{0.545098,0.482353,0.545098}
  \definecolor{gray0}{rgb}{0.000000,0.000000,0.000000}
  \definecolor{grey0}{rgb}{0.000000,0.000000,0.000000}
  \definecolor{gray1}{rgb}{0.011765,0.011765,0.011765}
  \definecolor{grey1}{rgb}{0.011765,0.011765,0.011765}
  \definecolor{gray2}{rgb}{0.019608,0.019608,0.019608}
  \definecolor{grey2}{rgb}{0.019608,0.019608,0.019608}
  \definecolor{gray3}{rgb}{0.031373,0.031373,0.031373}
  \definecolor{grey3}{rgb}{0.031373,0.031373,0.031373}
  \definecolor{gray4}{rgb}{0.039216,0.039216,0.039216}
  \definecolor{grey4}{rgb}{0.039216,0.039216,0.039216}
  \definecolor{gray5}{rgb}{0.050980,0.050980,0.050980}
  \definecolor{grey5}{rgb}{0.050980,0.050980,0.050980}
  \definecolor{gray6}{rgb}{0.058824,0.058824,0.058824}
  \definecolor{grey6}{rgb}{0.058824,0.058824,0.058824}
  \definecolor{gray7}{rgb}{0.070588,0.070588,0.070588}
  \definecolor{grey7}{rgb}{0.070588,0.070588,0.070588}
  \definecolor{gray8}{rgb}{0.078431,0.078431,0.078431}
  \definecolor{grey8}{rgb}{0.078431,0.078431,0.078431}
  \definecolor{gray9}{rgb}{0.090196,0.090196,0.090196}
  \definecolor{grey9}{rgb}{0.090196,0.090196,0.090196}
  \definecolor{gray10}{rgb}{0.101961,0.101961,0.101961}
  \definecolor{grey10}{rgb}{0.101961,0.101961,0.101961}
  \definecolor{gray11}{rgb}{0.109804,0.109804,0.109804}
  \definecolor{grey11}{rgb}{0.109804,0.109804,0.109804}
  \definecolor{gray12}{rgb}{0.121569,0.121569,0.121569}
  \definecolor{grey12}{rgb}{0.121569,0.121569,0.121569}
  \definecolor{gray13}{rgb}{0.129412,0.129412,0.129412}
  \definecolor{grey13}{rgb}{0.129412,0.129412,0.129412}
  \definecolor{gray14}{rgb}{0.141176,0.141176,0.141176}
  \definecolor{grey14}{rgb}{0.141176,0.141176,0.141176}
  \definecolor{gray15}{rgb}{0.149020,0.149020,0.149020}
  \definecolor{grey15}{rgb}{0.149020,0.149020,0.149020}
  \definecolor{gray16}{rgb}{0.160784,0.160784,0.160784}
  \definecolor{grey16}{rgb}{0.160784,0.160784,0.160784}
  \definecolor{gray17}{rgb}{0.168627,0.168627,0.168627}
  \definecolor{grey17}{rgb}{0.168627,0.168627,0.168627}
  \definecolor{gray18}{rgb}{0.180392,0.180392,0.180392}
  \definecolor{grey18}{rgb}{0.180392,0.180392,0.180392}
  \definecolor{gray19}{rgb}{0.188235,0.188235,0.188235}
  \definecolor{grey19}{rgb}{0.188235,0.188235,0.188235}
  \definecolor{gray20}{rgb}{0.200000,0.200000,0.200000}
  \definecolor{grey20}{rgb}{0.200000,0.200000,0.200000}
  \definecolor{gray21}{rgb}{0.211765,0.211765,0.211765}
  \definecolor{grey21}{rgb}{0.211765,0.211765,0.211765}
  \definecolor{gray22}{rgb}{0.219608,0.219608,0.219608}
  \definecolor{grey22}{rgb}{0.219608,0.219608,0.219608}
  \definecolor{gray23}{rgb}{0.231373,0.231373,0.231373}
  \definecolor{grey23}{rgb}{0.231373,0.231373,0.231373}
  \definecolor{gray24}{rgb}{0.239216,0.239216,0.239216}
  \definecolor{grey24}{rgb}{0.239216,0.239216,0.239216}
  \definecolor{gray25}{rgb}{0.250980,0.250980,0.250980}
  \definecolor{grey25}{rgb}{0.250980,0.250980,0.250980}
  \definecolor{gray26}{rgb}{0.258824,0.258824,0.258824}
  \definecolor{grey26}{rgb}{0.258824,0.258824,0.258824}
  \definecolor{gray27}{rgb}{0.270588,0.270588,0.270588}
  \definecolor{grey27}{rgb}{0.270588,0.270588,0.270588}
  \definecolor{gray28}{rgb}{0.278431,0.278431,0.278431}
  \definecolor{grey28}{rgb}{0.278431,0.278431,0.278431}
  \definecolor{gray29}{rgb}{0.290196,0.290196,0.290196}
  \definecolor{grey29}{rgb}{0.290196,0.290196,0.290196}
  \definecolor{gray30}{rgb}{0.301961,0.301961,0.301961}
  \definecolor{grey30}{rgb}{0.301961,0.301961,0.301961}
  \definecolor{gray31}{rgb}{0.309804,0.309804,0.309804}
  \definecolor{grey31}{rgb}{0.309804,0.309804,0.309804}
  \definecolor{gray32}{rgb}{0.321569,0.321569,0.321569}
  \definecolor{grey32}{rgb}{0.321569,0.321569,0.321569}
  \definecolor{gray33}{rgb}{0.329412,0.329412,0.329412}
  \definecolor{grey33}{rgb}{0.329412,0.329412,0.329412}
  \definecolor{gray34}{rgb}{0.341176,0.341176,0.341176}
  \definecolor{grey34}{rgb}{0.341176,0.341176,0.341176}
  \definecolor{gray35}{rgb}{0.349020,0.349020,0.349020}
  \definecolor{grey35}{rgb}{0.349020,0.349020,0.349020}
  \definecolor{gray36}{rgb}{0.360784,0.360784,0.360784}
  \definecolor{grey36}{rgb}{0.360784,0.360784,0.360784}
  \definecolor{gray37}{rgb}{0.368627,0.368627,0.368627}
  \definecolor{grey37}{rgb}{0.368627,0.368627,0.368627}
  \definecolor{gray38}{rgb}{0.380392,0.380392,0.380392}
  \definecolor{grey38}{rgb}{0.380392,0.380392,0.380392}
  \definecolor{gray39}{rgb}{0.388235,0.388235,0.388235}
  \definecolor{grey39}{rgb}{0.388235,0.388235,0.388235}
  \definecolor{gray40}{rgb}{0.400000,0.400000,0.400000}
  \definecolor{grey40}{rgb}{0.400000,0.400000,0.400000}
  \definecolor{gray41}{rgb}{0.411765,0.411765,0.411765}
  \definecolor{grey41}{rgb}{0.411765,0.411765,0.411765}
  \definecolor{gray42}{rgb}{0.419608,0.419608,0.419608}
  \definecolor{grey42}{rgb}{0.419608,0.419608,0.419608}
  \definecolor{gray43}{rgb}{0.431373,0.431373,0.431373}
  \definecolor{grey43}{rgb}{0.431373,0.431373,0.431373}
  \definecolor{gray44}{rgb}{0.439216,0.439216,0.439216}
  \definecolor{grey44}{rgb}{0.439216,0.439216,0.439216}
  \definecolor{gray45}{rgb}{0.450980,0.450980,0.450980}
  \definecolor{grey45}{rgb}{0.450980,0.450980,0.450980}
  \definecolor{gray46}{rgb}{0.458824,0.458824,0.458824}
  \definecolor{grey46}{rgb}{0.458824,0.458824,0.458824}
  \definecolor{gray47}{rgb}{0.470588,0.470588,0.470588}
  \definecolor{grey47}{rgb}{0.470588,0.470588,0.470588}
  \definecolor{gray48}{rgb}{0.478431,0.478431,0.478431}
  \definecolor{grey48}{rgb}{0.478431,0.478431,0.478431}
  \definecolor{gray49}{rgb}{0.490196,0.490196,0.490196}
  \definecolor{grey49}{rgb}{0.490196,0.490196,0.490196}
  \definecolor{gray50}{rgb}{0.498039,0.498039,0.498039}
  \definecolor{grey50}{rgb}{0.498039,0.498039,0.498039}
  \definecolor{gray51}{rgb}{0.509804,0.509804,0.509804}
  \definecolor{grey51}{rgb}{0.509804,0.509804,0.509804}
  \definecolor{gray52}{rgb}{0.521569,0.521569,0.521569}
  \definecolor{grey52}{rgb}{0.521569,0.521569,0.521569}
  \definecolor{gray53}{rgb}{0.529412,0.529412,0.529412}
  \definecolor{grey53}{rgb}{0.529412,0.529412,0.529412}
  \definecolor{gray54}{rgb}{0.541176,0.541176,0.541176}
  \definecolor{grey54}{rgb}{0.541176,0.541176,0.541176}
  \definecolor{gray55}{rgb}{0.549020,0.549020,0.549020}
  \definecolor{grey55}{rgb}{0.549020,0.549020,0.549020}
  \definecolor{gray56}{rgb}{0.560784,0.560784,0.560784}
  \definecolor{grey56}{rgb}{0.560784,0.560784,0.560784}
  \definecolor{gray57}{rgb}{0.568627,0.568627,0.568627}
  \definecolor{grey57}{rgb}{0.568627,0.568627,0.568627}
  \definecolor{gray58}{rgb}{0.580392,0.580392,0.580392}
  \definecolor{grey58}{rgb}{0.580392,0.580392,0.580392}
  \definecolor{gray59}{rgb}{0.588235,0.588235,0.588235}
  \definecolor{grey59}{rgb}{0.588235,0.588235,0.588235}
  \definecolor{gray60}{rgb}{0.600000,0.600000,0.600000}
  \definecolor{grey60}{rgb}{0.600000,0.600000,0.600000}
  \definecolor{gray61}{rgb}{0.611765,0.611765,0.611765}
  \definecolor{grey61}{rgb}{0.611765,0.611765,0.611765}
  \definecolor{gray62}{rgb}{0.619608,0.619608,0.619608}
  \definecolor{grey62}{rgb}{0.619608,0.619608,0.619608}
  \definecolor{gray63}{rgb}{0.631373,0.631373,0.631373}
  \definecolor{grey63}{rgb}{0.631373,0.631373,0.631373}
  \definecolor{gray64}{rgb}{0.639216,0.639216,0.639216}
  \definecolor{grey64}{rgb}{0.639216,0.639216,0.639216}
  \definecolor{gray65}{rgb}{0.650980,0.650980,0.650980}
  \definecolor{grey65}{rgb}{0.650980,0.650980,0.650980}
  \definecolor{gray66}{rgb}{0.658824,0.658824,0.658824}
  \definecolor{grey66}{rgb}{0.658824,0.658824,0.658824}
  \definecolor{gray67}{rgb}{0.670588,0.670588,0.670588}
  \definecolor{grey67}{rgb}{0.670588,0.670588,0.670588}
  \definecolor{gray68}{rgb}{0.678431,0.678431,0.678431}
  \definecolor{grey68}{rgb}{0.678431,0.678431,0.678431}
  \definecolor{gray69}{rgb}{0.690196,0.690196,0.690196}
  \definecolor{grey69}{rgb}{0.690196,0.690196,0.690196}
  \definecolor{gray70}{rgb}{0.701961,0.701961,0.701961}
  \definecolor{grey70}{rgb}{0.701961,0.701961,0.701961}
  \definecolor{gray71}{rgb}{0.709804,0.709804,0.709804}
  \definecolor{grey71}{rgb}{0.709804,0.709804,0.709804}
  \definecolor{gray72}{rgb}{0.721569,0.721569,0.721569}
  \definecolor{grey72}{rgb}{0.721569,0.721569,0.721569}
  \definecolor{gray73}{rgb}{0.729412,0.729412,0.729412}
  \definecolor{grey73}{rgb}{0.729412,0.729412,0.729412}
  \definecolor{gray74}{rgb}{0.741176,0.741176,0.741176}
  \definecolor{grey74}{rgb}{0.741176,0.741176,0.741176}
  \definecolor{gray75}{rgb}{0.749020,0.749020,0.749020}
  \definecolor{grey75}{rgb}{0.749020,0.749020,0.749020}
  \definecolor{gray76}{rgb}{0.760784,0.760784,0.760784}
  \definecolor{grey76}{rgb}{0.760784,0.760784,0.760784}
  \definecolor{gray77}{rgb}{0.768627,0.768627,0.768627}
  \definecolor{grey77}{rgb}{0.768627,0.768627,0.768627}
  \definecolor{gray78}{rgb}{0.780392,0.780392,0.780392}
  \definecolor{grey78}{rgb}{0.780392,0.780392,0.780392}
  \definecolor{gray79}{rgb}{0.788235,0.788235,0.788235}
  \definecolor{grey79}{rgb}{0.788235,0.788235,0.788235}
  \definecolor{gray80}{rgb}{0.800000,0.800000,0.800000}
  \definecolor{grey80}{rgb}{0.800000,0.800000,0.800000}
  \definecolor{gray81}{rgb}{0.811765,0.811765,0.811765}
  \definecolor{grey81}{rgb}{0.811765,0.811765,0.811765}
  \definecolor{gray82}{rgb}{0.819608,0.819608,0.819608}
  \definecolor{grey82}{rgb}{0.819608,0.819608,0.819608}
  \definecolor{gray83}{rgb}{0.831373,0.831373,0.831373}
  \definecolor{grey83}{rgb}{0.831373,0.831373,0.831373}
  \definecolor{gray84}{rgb}{0.839216,0.839216,0.839216}
  \definecolor{grey84}{rgb}{0.839216,0.839216,0.839216}
  \definecolor{gray85}{rgb}{0.850980,0.850980,0.850980}
  \definecolor{grey85}{rgb}{0.850980,0.850980,0.850980}
  \definecolor{gray86}{rgb}{0.858824,0.858824,0.858824}
  \definecolor{grey86}{rgb}{0.858824,0.858824,0.858824}
  \definecolor{gray87}{rgb}{0.870588,0.870588,0.870588}
  \definecolor{grey87}{rgb}{0.870588,0.870588,0.870588}
  \definecolor{gray88}{rgb}{0.878431,0.878431,0.878431}
  \definecolor{grey88}{rgb}{0.878431,0.878431,0.878431}
  \definecolor{gray89}{rgb}{0.890196,0.890196,0.890196}
  \definecolor{grey89}{rgb}{0.890196,0.890196,0.890196}
  \definecolor{gray90}{rgb}{0.898039,0.898039,0.898039}
  \definecolor{grey90}{rgb}{0.898039,0.898039,0.898039}
  \definecolor{gray91}{rgb}{0.909804,0.909804,0.909804}
  \definecolor{grey91}{rgb}{0.909804,0.909804,0.909804}
  \definecolor{gray92}{rgb}{0.921569,0.921569,0.921569}
  \definecolor{grey92}{rgb}{0.921569,0.921569,0.921569}
  \definecolor{gray93}{rgb}{0.929412,0.929412,0.929412}
  \definecolor{grey93}{rgb}{0.929412,0.929412,0.929412}
  \definecolor{gray94}{rgb}{0.941176,0.941176,0.941176}
  \definecolor{grey94}{rgb}{0.941176,0.941176,0.941176}
  \definecolor{gray95}{rgb}{0.949020,0.949020,0.949020}
  \definecolor{grey95}{rgb}{0.949020,0.949020,0.949020}
  \definecolor{gray96}{rgb}{0.960784,0.960784,0.960784}
  \definecolor{grey96}{rgb}{0.960784,0.960784,0.960784}
  \definecolor{gray97}{rgb}{0.968627,0.968627,0.968627}
  \definecolor{grey97}{rgb}{0.968627,0.968627,0.968627}
  \definecolor{gray98}{rgb}{0.980392,0.980392,0.980392}
  \definecolor{grey98}{rgb}{0.980392,0.980392,0.980392}
  \definecolor{gray99}{rgb}{0.988235,0.988235,0.988235}
  \definecolor{grey99}{rgb}{0.988235,0.988235,0.988235}
  \definecolor{gray100}{rgb}{1.000000,1.000000,1.000000}
  \definecolor{grey100}{rgb}{1.000000,1.000000,1.000000}
  \definecolor{dark grey}{rgb}{0.662745,0.662745,0.662745}
  \definecolor{DarkGrey}{rgb}{0.662745,0.662745,0.662745}
  \definecolor{dark gray}{rgb}{0.662745,0.662745,0.662745}
  \definecolor{DarkGray}{rgb}{0.662745,0.662745,0.662745}
  \definecolor{dark blue}{rgb}{0.000000,0.000000,0.545098}
  \definecolor{DarkBlue}{rgb}{0.000000,0.000000,0.545098}
  \definecolor{dark cyan}{rgb}{0.000000,0.545098,0.545098}
  \definecolor{DarkCyan}{rgb}{0.000000,0.545098,0.545098}
  \definecolor{dark magenta}{rgb}{0.545098,0.000000,0.545098}
  \definecolor{DarkMagenta}{rgb}{0.545098,0.000000,0.545098}
  \definecolor{dark red}{rgb}{0.545098,0.000000,0.000000}
  \definecolor{DarkRed}{rgb}{0.545098,0.000000,0.000000}
  \definecolor{light green}{rgb}{0.564706,0.933333,0.564706}
  \definecolor{LightGreen}{rgb}{0.564706,0.933333,0.564706}